\documentclass[aps,pre,floats,twocolumn,superscriptaddress,floatfix]{revtex4-1}
\usepackage{amsmath,amsfonts,amssymb,subfigure,multirow}
\usepackage[pdftex]{graphicx}

\begin{document}

\title{How Small Are Building Blocks of Complex Networks}

\author{Almerima Jamakovic}
\affiliation{TNO Information and Communication Technology, Netherlands Organisation for Applied Scientific Research, P.O. Box 5050, 2600 GB Delft, The Netherlands}

\author{Priya Mahadevan}
\affiliation{HP Labs, 1501 Page Mill Rd, Palo Alto, California, 94304, USA}

\author{Amin Vahdat}
\affiliation{Department of Computer Science and Engineering, University of California, San Diego (UCSD), 9500 Gilman Drive, La Jolla, California 92093, USA}

\author{Mari{\'a}n Bogu{\~n}{\'a}}
\affiliation{Departament de F{\'\i}sica Fonamental, Universitat de Barcelona, Mart\'{\i} i Franqu\`es 1, 08028 Barcelona, Spain}

\author{Dmitri Krioukov}
\affiliation{Cooperative Association for Internet Data Analysis (CAIDA), University of California, San Diego (UCSD), 9500 Gilman Drive, La Jolla, California 92093, USA}

\begin{abstract}
Network motifs are small building blocks of complex networks.
Statistically significant motifs often perform network-specific
functions. However, the precise nature of the connection between
motifs and the global structure and function of networks remains
elusive. Here we show that the global structure of some real
networks is statistically determined by the probability of
connections within motifs of size at most~$3$, once this probability
accounts for node degrees. The connectivity profiles of node triples
in these networks capture all their local and global properties.
This finding impacts methods relying on motif statistical
significance, and enriches our understanding of the elementary
forces that shape the structure of complex networks.
\end{abstract}

\maketitle

\section{Introduction}

A promising direction in the study of the structure and function
of complex networks is to identify their building blocks, or
motifs~\cite{MiSh02-motifs,Alon07-review,Alon06-book}, which are
small subgraphs in a real network.
A great deal of research---in particular, research on gene
regulatory networks---shows that specific motifs perform specific
functions, such as speeding up response times of regulatory
networks~\cite{RoEl02,MaIt06}.
However, motifs have also raised many
questions~\cite{KnNe08,InSt06,CoHo06,KuBa06,MaBo05,SaKo05,VaDo04,ArRa04},
including continuing debates on whether and how motif statistical profiles are related to the {\em global\/} structure,
function, and evolution of certain networks.
Our recent work~\cite{MaKrFaVa06-phys} introduces $dK$-series,
see Section~\ref{sec:dk-series}.

The $dK$-series, with analogy to the
Taylor or Fourier series, is a systematic and complete basis
for characterizing network structure. The $dK$-series subsumes
many known motif- and degree-based statistical characteristics of
complex networks.
While motifs are subgraphs whose nodes can have any degree in a given
real network, the $dK$-series preserves the information about these
degrees. For example, the zero-th element of the $dK$-series, the
$0K$-``distribution,'' is the average degree in a given network. The
first element, the $1K$-distribution, is the network's degree distribution,
or the number of nodes---subgraphs of size $1$---of
degree $k$. The second element, the $2K$-distribution, is the joint
degree distribution, the number of subgraphs of size
$2$---links---involving nodes of degrees $k_1$ and $k_2$. The $2K$-distribution
thus defines $2$-node degree correlations. For $d=3$,
the subgraphs are triangles and wedges, composed of nodes of degrees
$k_1$, $k_2$, and $k_3$, which defines clustering. Generalizing, the $dK$-distribution is the
numbers of different subgraphs of size $d$ involving nodes of degrees
$k_1, k_2, \ldots, k_d$.

The $dK$-series is systematic and complete because it is {\em inclusive}
and {\em converging}. Inclusiveness results from the fact
that the $(d+1)K$-distribution contains the same information about the
network as the $dK$-distribution, plus some additional
information. That is, by increasing $d$, we provide strictly more
detail about the network structure, which is not the case with a
motif-based series. Node degrees are necessary to make the series
inclusive and thus systematic,
see Section~\ref{sec:motifs-vs-dk}.
As $d$ increases toward the network
size, we fully specify the entire network structure, which
explains the second {\em convergence\/} property of $dK$-series---it
converges to the given network in the limit of large $d$.

Does this convergence happen only at $d$ equal to the network size,
or much sooner, at smaller $d$? In other words, how much {\em
local\/} information, i.e., information about concentrations of
degree-labeled subgraphs of what size, is needed to fully
capture {\em global\/} network structure?

\begin{figure*}
\centerline{\includegraphics[width=\linewidth]{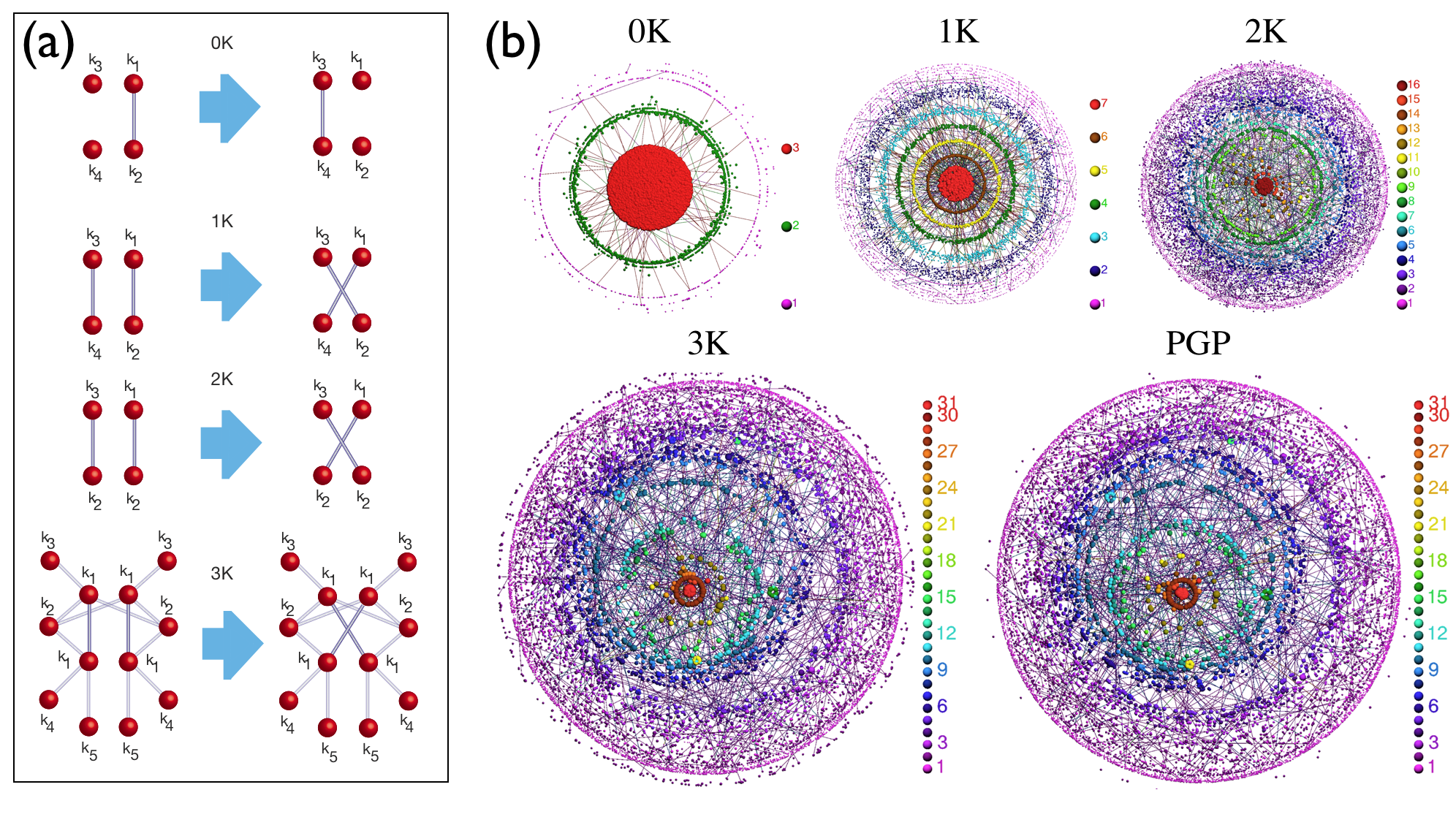}}
\caption{{\bf The $dK$-randomization null models for $d=0,1,2,3$.}
{\bf a)~Illustration of $dK$-randomizing rewiring.} All nodes are
labeled by their degrees, and a $dK$-rewiring preserves the graph's
$dK$-distribution, and consequently its $d'K$-distributions for all
$d'<d$, but randomizes the $d''K$-distributions for $d''>d$. The
$0K$-randomization involves rewiring of a link to any pair of
disconnected nodes, which preserves the average degree only. The
$1K$-randomization preserves the degree distribution, too, by
rewiring a pair of links as shown. The $2K$-distribution preserves
the joint degree distribution as well, because at least two nodes
adjacent to the rewired links are of the same degree. The
$3K$-randomization preserves the number of degree-labeled wedges and
triangles. As $d$ increases, the rewiring becomes increasingly more
constrained since fewer links can be rewired without altering the
$dK$-distribution. To $dK$-randomize a network, we randomly select a
pair of links, and rewire them if they can be $dK$-rewired, or, if
they cannot be rewired, select another random pair. This process is
repeated for a sufficient number of successful rewirings, i.e.,
until all network properties stop changing, at which point we say
that the graph has converged to its $dK$-randomization. {\bf
b)~Visualization of the social web of trust (PGP
network~\cite{BoPa04}) and its $dK$-randomizations.} We use the
LaNet-vi tool~\cite{Lanet-vi} for visualization, which encodes the
node coreness in color, see the right legends. The coreness is a
measure of node centrality, i.e., how
deeply in the network core the node lies~\cite{Lanet-vi}. Nodes with
larger coreness are also placed closer to the circle centers. The quick
convergence of the $dK$-randomizations to the original PGP network,
and the similarity between it and its $3K$-randomization are
remarkable. \label{fig:dk-series}}
\end{figure*}

To answer these questions, we must compare a real network with
typical random networks defined by its $dK$-distribution. If there
is no difference between such {\em $dK$-random\/} networks and the
real network, then the latter is fixed by its $dK$-distribution. To
obtain a $dK$-random version of the real network, we $dK$-randomize
it as illustrated in Fig.~\ref{fig:dk-series}(a)---we randomly
rewire (pairs of) links preserving the $dK$-distribution in the
network, generalizing known network randomization
techniques~\cite{MaSne02,MaSne03} used to compute motif statistical
significance. The result of this $dK$-randomization procedure are
random networks that have the same $dK$-distribution as the original
real network, but that are maximally random in all other respects.

Our question thus becomes what is the minimum value of $d$ such that
there is no difference between a real network and its
$dK$-randomizations? It seems at first that the answer to this
question should strongly depend on the specific networks we
consider.

We consider a variety of social, biological, transportation,
communication, and technological networks,
see Section~\ref{sec:real-networks}.
Although the $dK$-series applies to directed and even
annotated networks~\cite{DiKr09},
here we report results for undirected networks. The
$dK$-distributions for directed or annotated networks contain more
information than for undirected networks. Therefore, $dK$-series
converges faster in the former case~\cite{DiKr09}. Below we show the
results for the well-studied social web of trust relationships
extracted from Pretty Good Privacy (PGP) data~\cite{BoPa04}. We
obtain similar results for all other considered networks (protein
interactions in {\it Saccharomyces cerevisiae}, scientific collaborations,
US air transportation, and the Internet), except for the power grid,
see~Section~\ref{sec:dk-comparisons},
where we also discuss possible reasons
for why the power grid appears as an exception.

Fig.~\ref{fig:dk-series}(b) visualizes the PGP network and its
$dK$-randomizations. We observe that the $dK$-series
converges at $d=3$. While the $0K$-random network has little in
common with the real network, the $1K$-random one is somewhat more
similar, even more so for $2K$, and there is very little difference
between the real PGP network and its $3K$-random counterpart.

To provide a more detailed and insightful comparison between the
real network and its $dK$-randomizations,
we compute a variety of metrics for each.
Some popular metrics, such as degree distribution, average nearest neighbor
connectivity, clustering, etc., are functions, sometimes peculiar,
of $dK$-distributions, and therefore it is not surprising that they
are properly captured by $dK$-series, as we confirm in
Section~\ref{sec:dk-metrics}.
We classify metrics that do not explicitly depend on
$dK$-distributions as microscopic,
mesoscopic, and macroscopic. We choose them to probe the network
structure at the local, medium, and global scales.

\begin{figure}
\centerline{\includegraphics[width=\linewidth]{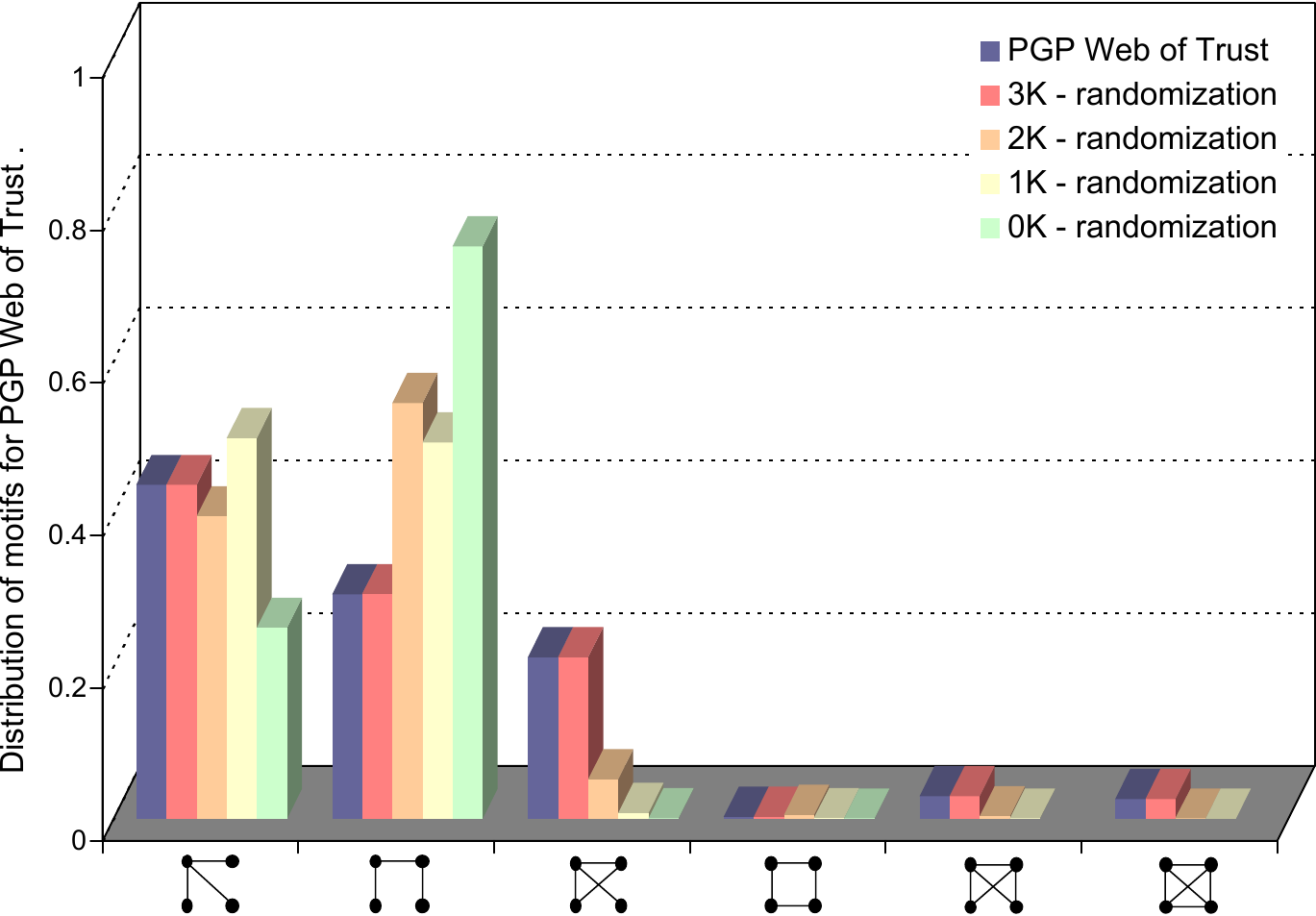}}
\centerline{\includegraphics[width=\linewidth]{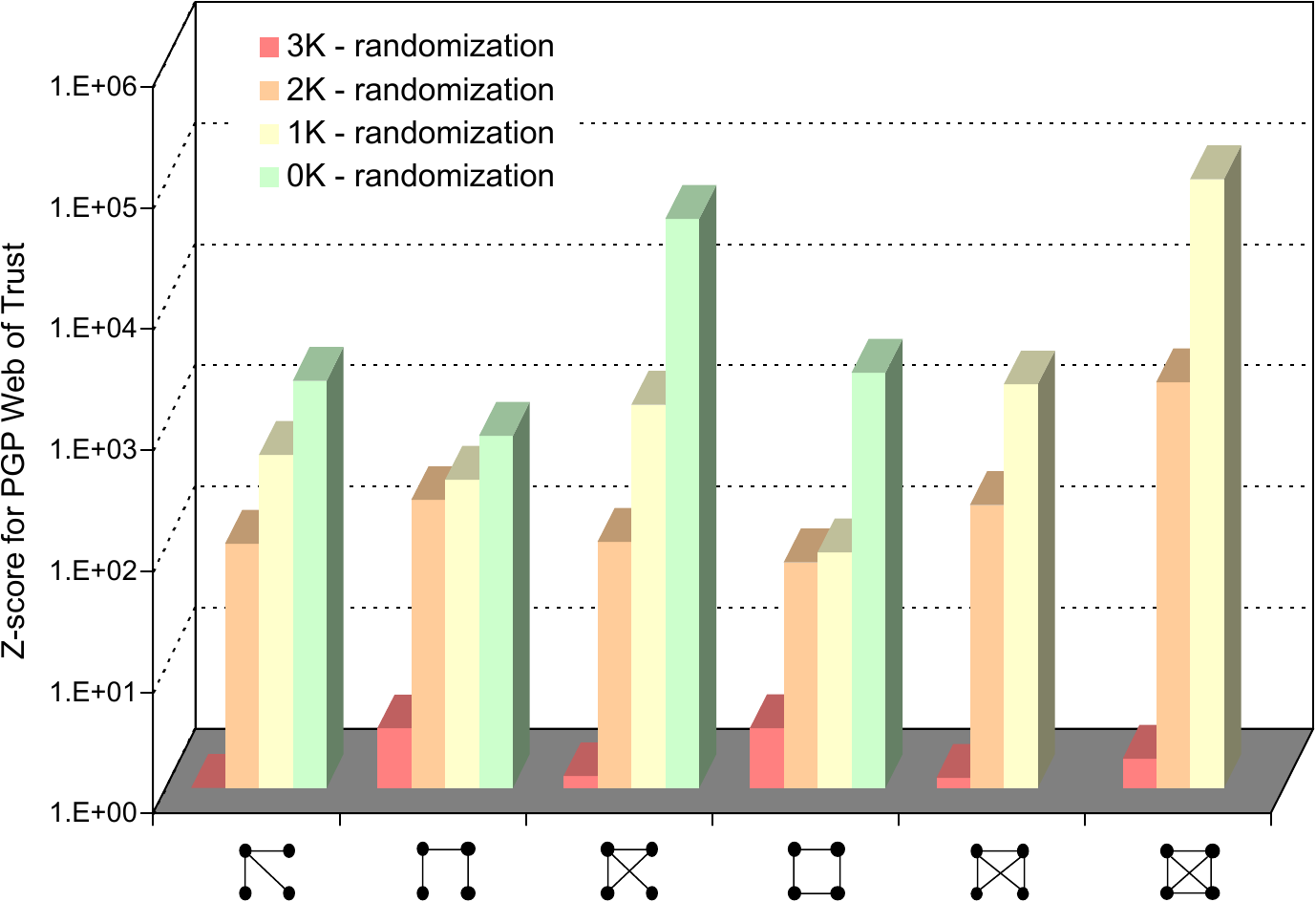}}
\caption{{\bf Microscopic scale: motifs.} There are six different
graphs of size $4$ shown on the $x$-axes. The
{\bf left}
plot shows the distribution of the numbers of these subgraphs in the
PGP network and its $dK$-randomizations, $d=0,1,2,3$. Each blue bar,
for example, is the number of the corresponding subgraph occurrences
in the PGP network divided by the total number of subgraphs of size
$4$ in it. For $dK$-randomizations, the values are averaged, for
each $d$, over several realizations of the $dK$-randomized network.
In the case of $0K$-randomization, the last two motifs did not occur
in any randomized sample of the network. The
{\bf right}
plot shows the Z-scores for the six subgraphs in the four
$dK$-randomization null models. The Z-score~\cite{MiSh02-motifs} of
a subgraph is a measure of its statistical significance in a real
network, compared to a randomization null model. Specifically, the
Z-score $Z$ is the difference between the number $N$ of the
occurrences of a subgraph in the real network and the average number
$\bar{N}$ of its occurrences in the corresponding randomized
networks, divided by the standard deviation $\sigma$ of its
occurrences in the randomized networks,
$Z=|N-\bar{N}|/\sigma$.\label{fig:micro-properties}}
\end{figure}

The simplest {\em microscopic}, local-structure statistics, which
are not fixed by the $dK$-distributions with $d\leq3$, are the
frequencies of motifs of size $4$ without degree information.
We compute these frequencies in
the real network and its $dK$-randomizations, and show the results
in Fig.~\ref{fig:micro-properties}. We find that
no motif is statistically significant for $d=3$,
i.e., all the Z-scores are small, and all the
motif frequencies in the real network and its $3K$-randomization
are virtually the same.

\begin{figure}
\centerline{\includegraphics[width=\linewidth]{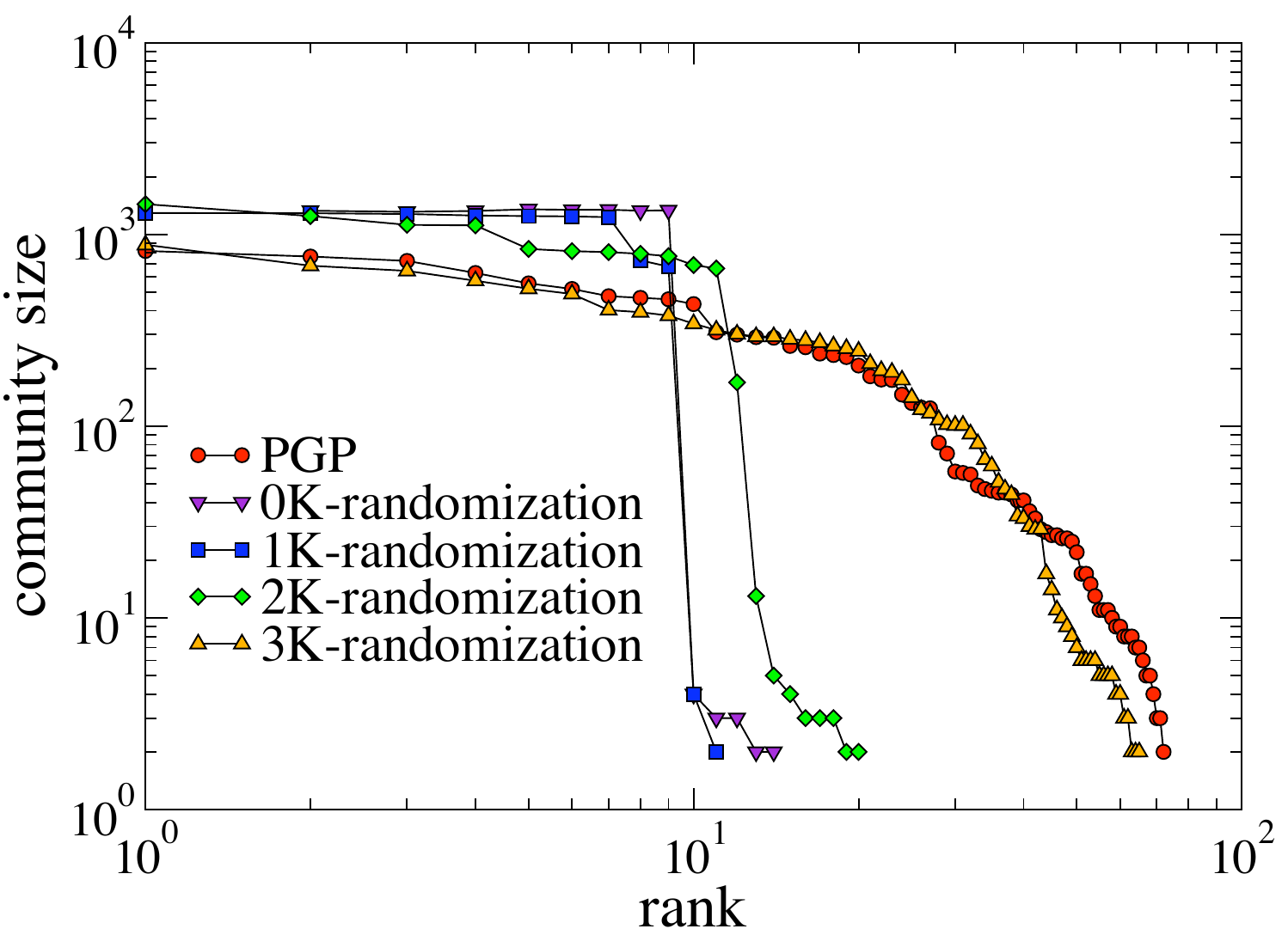}}
\caption{{\bf Mesoscopic scale: community structure.} We compute
communities in the PGP network using the Extremal Optimization
algorithm~\cite{DuAr05}. We then sort the found communities in the
order of decreasing size. The size of a community is the number of
nodes in it. The rank of a community is its position number in the
size-ordered list. We then show the community size distribution by
plotting the community sizes vs.\ their ranks. \label{fig:meso-properties}}
\end{figure}

At the {\em mesoscopic\/} scale, we consider the community structure
of the PGP network. A community is a subgraph with many internal
connections, and a relatively small number of connections external to
the subgraph. Fig.~\ref{fig:meso-properties} shows that the
community structure is indeed a ``mesoscopic'' metric because the
community sizes range from a few nodes to
thousands of nodes for largest communities.
Fig.~\ref{fig:meso-properties} shows that the community size
distributions in the PGP network and its $3K$-randomization are very
similar.

\begin{figure}
\centerline{\includegraphics[width=\linewidth]{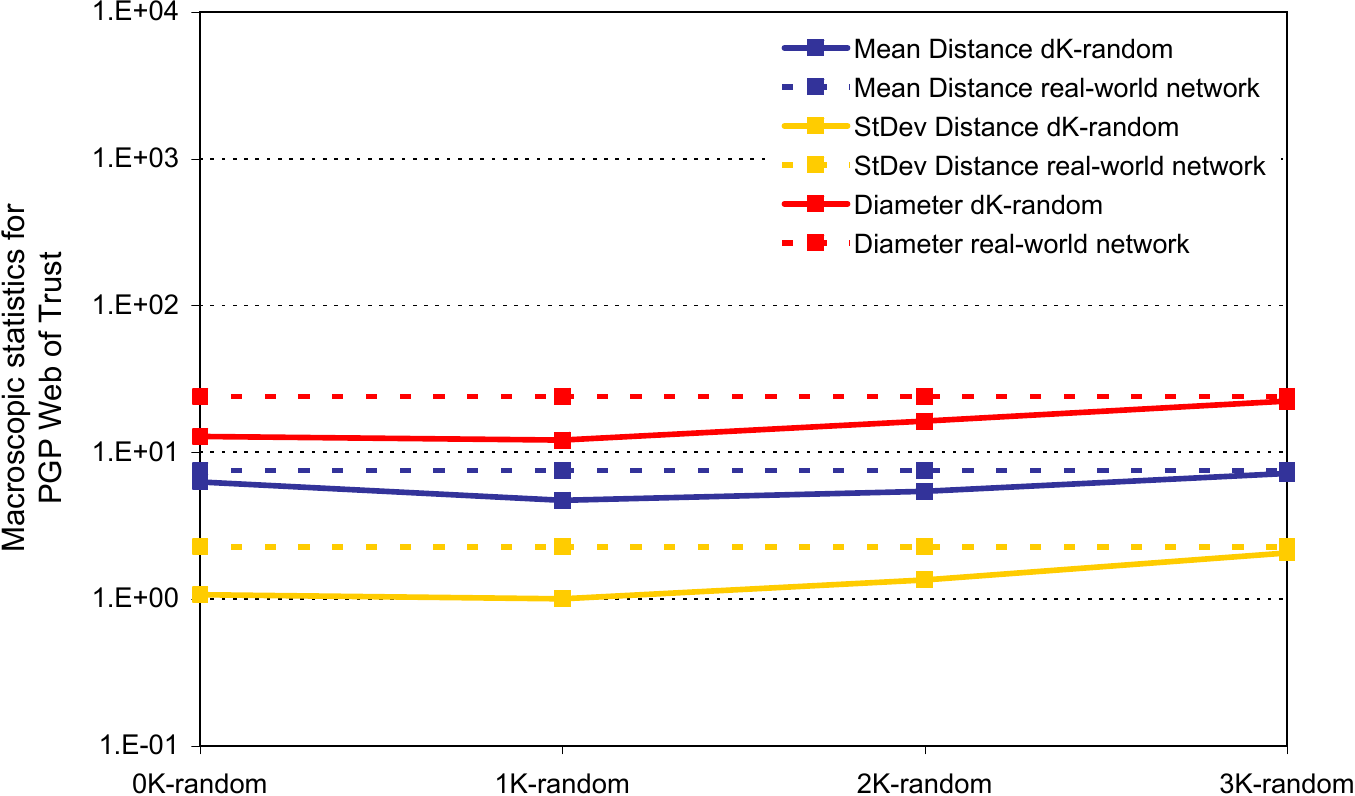}}
\centerline{\includegraphics[width=\linewidth]{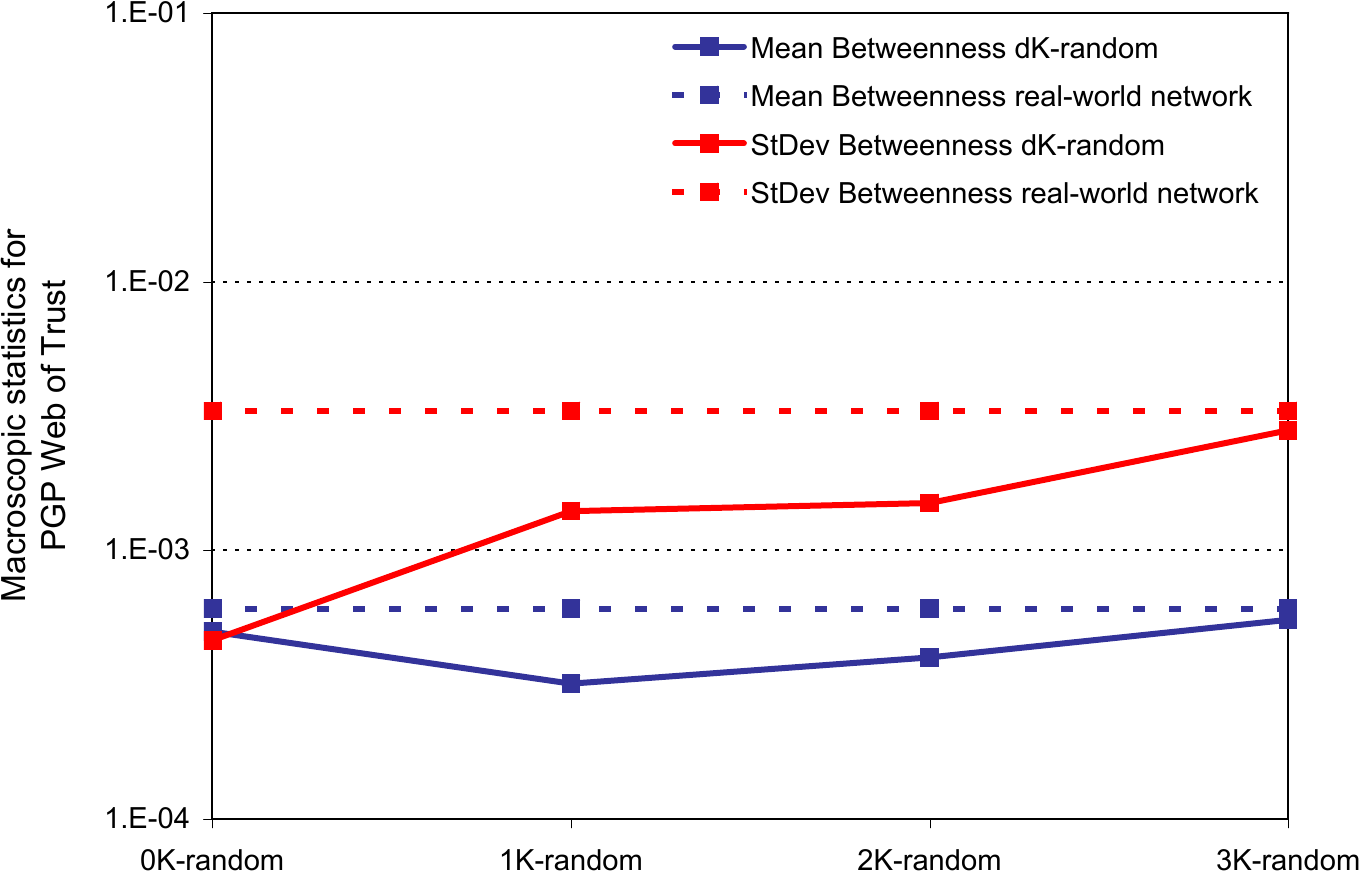}}
\caption{{\bf Macroscopic scale: the distance and betweenness
distributions.} The
{\bf left}
plot shows the metrics related to the hop length of shortest paths,
or distances, between nodes in the PGP network and its
$dK$-randomizations. These metrics are the average and maximum
distance between nodes, the latter called the network {\em
diameter}, and the standard deviation of the distance distribution.
The
{\bf right}
plot shows the average betweenness and the standard deviation of the
betweenness distribution of nodes in the PGP network and its
$dK$-randomizations. The betweenness of a node is a measure of its
communication centrality~\cite{Freeman77}. It is equal to the number
of shortest paths passing through the node, divided by the total
number of shortest paths between the same source and destination,
summed over all source-destinations pairs. In both plots the values
for $dK$-randomizations are averaged, for each $d$, over several
realizations of the $dK$-randomized network.
\label{fig:macro-properties}}
\end{figure}

At the {\em macroscopic} scale, we consider two popular and
important statistics that depend on a network's global structure: the
node betweenness centrality and the distribution of lengths of
shortest paths in a network. Fig.~\ref{fig:macro-properties} once
again shows that $3K$ is sufficient to capture even such global graph
properties; the considered metrics are approximately the same for the
PGP network and its $3K$-randomization.

We call a given real network \emph{$dK$-random} if {\em all\/} its
metrics, at all scales from local to global, are approximately the
same as the corresponding metrics in its $dK$-randomizations. We see
in Section~\ref{sec:dk-comparisons}
that in agreement with the results of
V\'azquez {\it et al.}~\cite{VaDo04}, almost all networks that we
collected data for are $3K$-random at most (some networks are $2K$-
or even $1K$-random). In other words, {\em the global structure of
these networks is captured entirely by the distribution of node
triples and their degrees}.

It is an open question why many different real networks are
$3K$-random. A trivial answer would be that $d=3$ is just
``constraining enough.''  There may only be a few possible rewirings
preserving the $3K$-distribution. But why exactly is $d=3$ sufficient
for real networks? There are many classes of synthetic graphs, such as
latices, for which no $d$ substantially smaller than the graph size is
``constraining enough.'' Perhaps the answer can be obtained by
studying the hidden metric spaces underlying real
networks~\cite{BoKrKc08}. The distances in such spaces abstract
intrinsic similarities between nodes. If these spaces are metric---and
we have found empirical evidence that they are indeed
such~\cite{SeKrBo08}---then the triangle inequality naturally yields
and explains network clustering, which the $3K$-distribution captures
by definition.

Regardless of the actual explanation, our results have diverse
implications. First, our $dK$-randomization basis makes it clear
that there is no {\it a priori\/} preferred null model for network randomization.
To tell how statistically important a given motif is, it is necessary to
compare its frequency in the real network with its frequency in
a network randomization, a null model. But one can $dK$-randomize any
network for any $d$, and we find that the (relative)
statistical significance of motifs strongly depends on $d$ for all
the considered networks. Therefore choosing any specific value of $d$,
or more generally, any specific null model
to compute motif significance requires some non-trivial justification.

Second, our finding that many networks are $3K$-random can assist our
understanding of how functions of an evolving network shape its
structure. Indeed, one can potentially simplify such explanations to
how the observed $3K$-distribution has emerged in the network. As soon
as one explains the emergence of the $3K$-distribution, all other
structural properties of the network follow as a consequence of its $3K$-randomness.

Finally, our work has practical implications for the design of
network topology models and generators. Many scientific disciplines,
including biology~\cite{KuBa06,BuLe06,KnNe08a,RoWe08} and computer
science~\cite{Waxman88,ZeCaBh96-phys,MeLaMaBy01-phys,WiJa02},
require laboratory modeling of real networks, and in particular the
ability to generate random graphs that reproduce their important
structural properties. Experimentation with the $dK$-series
representation of real networks has revealed its theoretical and
practical strengths for this task---one can reproduce all important
structural properties of a given network by generating its
$3K$-random graphs. Recent work~\cite{DiKr09} extending the
$dK$-series construction to support rich semantic, structural, or
functional annotations of nodes and links in real networks suggests
a considerable untapped potential of $dK$-series for many
disciplines.

\section{The $dK$-series illustrated}\label{sec:dk-series}

\begin{figure*}
\centerline{\includegraphics[width=6.5in]{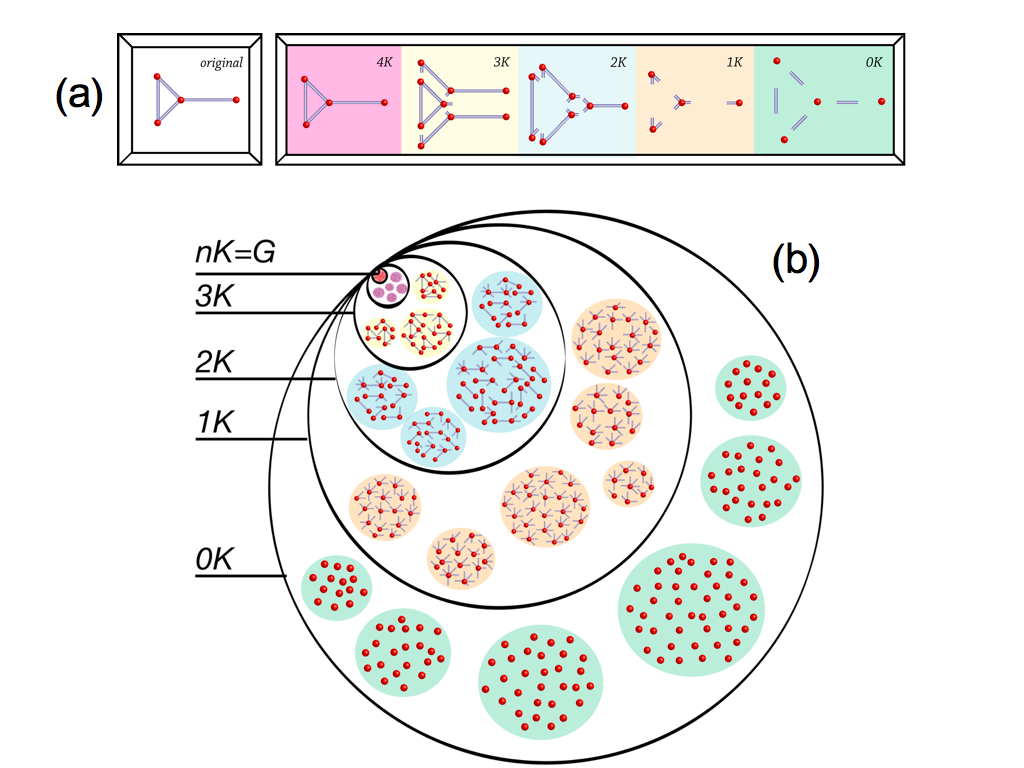}}
\caption{The $dK$-series illustrated: {\bf a)}~$dK$-distributions
for a graph of size $4$; {\bf b)}~convergence and inclusiveness of
$dK$-series. \label{fig:dk-series-appendix}}
\end{figure*}

In Fig.~\ref{fig:dk-series-appendix}(a) we illustrate $dK$-series
for a graph of size $4$. The $4K$-distribution is the graph itself.
The $3K$-distribution consists of its three subgraphs of size $3$:
one triangle connecting nodes of degrees $2$, $2$, and $3$, and two
wedges connecting nodes of degrees $2$, $3$, and $1$. The
$2K$-distribution is the joint degree distribution in the graph. It
specifies the number of links (subgraphs of size $2$) connecting
nodes of different degrees: one link connects nodes of degrees $2$
and $2$, two links connect nodes of degrees $2$ and $3$, and one
link connects nodes of degree $3$ and $1$. The $1K$-distribution is
the degree distribution in the graph. It lists the number of nodes
(subgraphs of size $1$) of different degree: one node of degree $1$,
two nodes of degree $2$, and one node of degree $3$. The
$0K$-distribution is just the average degree in the graph, which is
$2$.

Fig.~\ref{fig:dk-series-appendix}(b) illustrates the inclusiveness
and convergence of $dK$-series by showing the hierarchy of
$dK$-graphs, which are graphs that have the same $dK$-distribution
as some graph $G$ of size $n$. The black circles schematically shows
the sets of $dK$-graphs.

The set of $0K$-graphs is largest: the number of different graphs
that have the same average degree as $G$ is enormous. These graphs
may have a structure drastically different from $G$'s. The set of
$1K$-graphs is a subset of $0K$-graphs, because each graph with the
same degree distribution as in $G$ has also the same average degree
as $G$, but not {\it vice versa}. As a consequence, typical
(``maximally random'') $1K$-graphs tend to be more similar to $G$
than $0K$-graphs. The set of $2K$-graphs is a subset of $1K$-graphs,
also containing $G$.

As $d$ increases, the circles become smaller because the number of
different $dK$-graphs decreases. Since all the $dK$-graph sets
contain $G$, the circles ``zoom-in'' on it, and while their number
decreases, $dK$-graphs become increasingly more similar to $G$. In
the $d=n$ limit, the set of $nK$-graphs consists of only one
element, $G$ itself.

\section{The real networks considered}\label{sec:real-networks}

We collected data for a number of real networks. We wanted the set
of considered networks to be representative, in the sense that it
should contain networks of different nature, coming from different
domains, thus showing the universality of our $dK$-basis. The
considered networks include social, biological, transportation, and
technological networks. Specifically, we report results for:
\begin{itemize}
\item
The social web of trust relationships among people. The trust
relationships are inferred using the data from the Pretty Good
Privacy (PGP) encryption algorithm~\cite{BoPa04}. We extract the
strongly connected component from this network. The nodes are
people, and there is a link between two people if they trust each
other.
\item
The social network of scientific collaborations extracted from the
{\tt arXiv} condensed-matter database~\cite{newman01}. The nodes are
authors, and there is a link between two authors if they co-authored
a paper.
\item
The biological network of protein interactions in the yeast {\it
Saccharomyces cerevisiae} collected from the database of interacting
proteins~\cite{CoFlSeVe06}. The nodes are proteins, and there is a
link between two proteins if they interact.
\item
The US air transportation network~\cite{CoPaSaVe07}. The nodes are
airports, and there is a link between two airports if there is a
direct flight between them.
\item
The topology of the Internet at the level of Autonomous
Systems~(ASs)~\cite{MaKrFo06}. The nodes are ASs, i.e.,
organizations owing parts of the Internet infrastructure, and there
is a link between two ASs if they are physically connected.
\item
The electrical power grid in the western US~\cite{WatStr98}. The
nodes are generators, transformers, or substations, two of which are
linked if there is a high-voltage transmission line between them.
\end{itemize}
Table~\ref{table real-world networks} lists these
networks and their abbreviations used in the subsequent figures and
tables.

\begin{table}
\caption{The considered networks and their abbreviations.}
\label{table real-world networks}
\vspace{0.5cm}
\begin{tabular}{|l|l|}
\hline Network & Abbreviation
\\ \hline \hline PGP Web of Trust~\cite{BoPa04} & PGP
\\ \hline Scientific collaboration network~\cite{newman01} & Collab.
\\ \hline Protein interaction network~\cite{CoFlSeVe06} & Protein
\\ \hline US air transportation network~\cite{CoPaSaVe07} & Air
\\ \hline Internet at the level of ASs~\cite{MaKrFo06} & Internet
\\ \hline Power grid in the western US~\cite{WatStr98} & Power
\\ \hline
\end{tabular}
\end{table}

\section{Topologies of real networks and their $dK$-randomizations}\label{sec:dk-comparisons}

In this section we compare the real networks to their
$dK$-randomizations across a number of topological metrics.

\begin{figure*}
\centerline{\includegraphics[width=.9\linewidth]{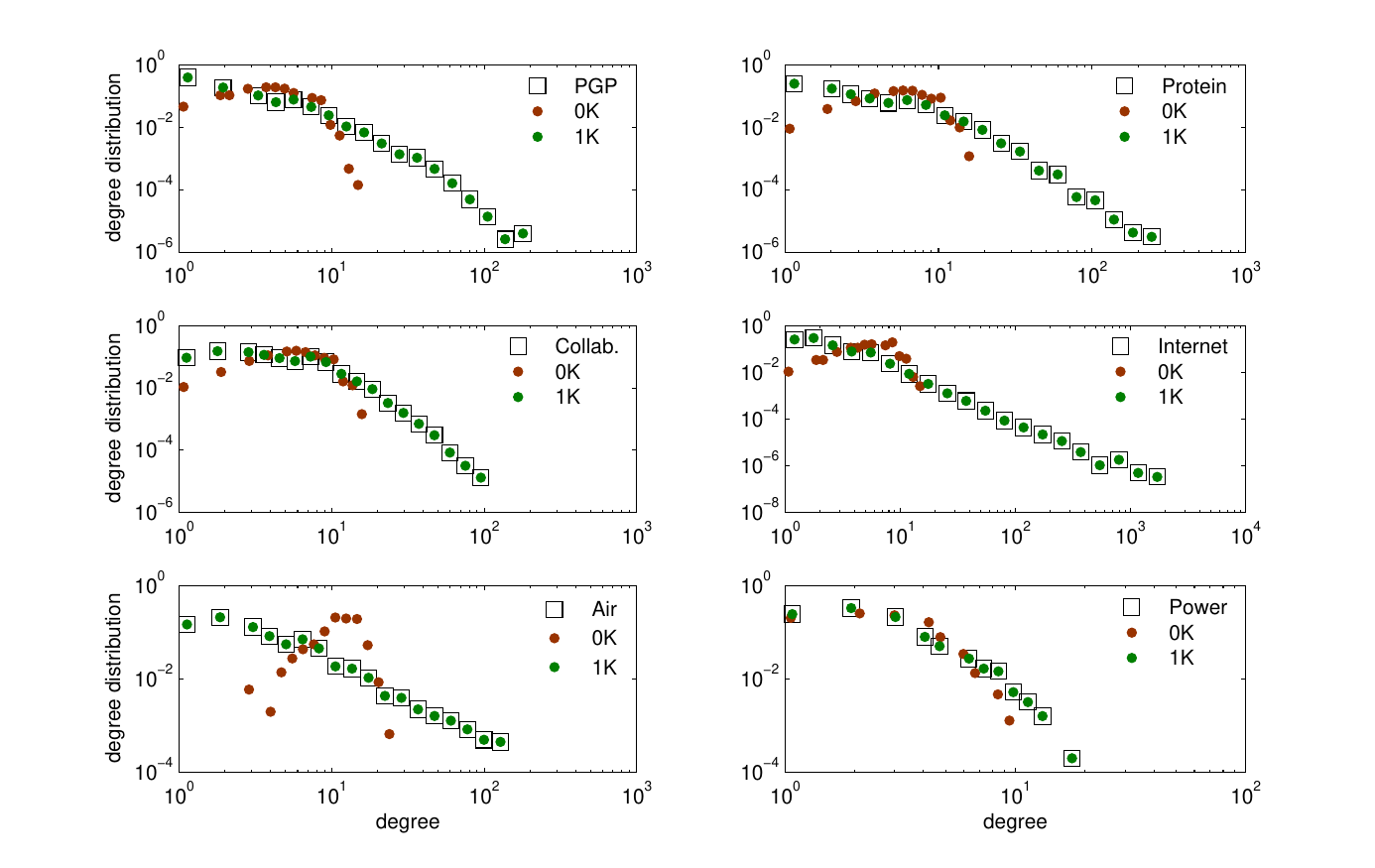}}
\caption{The degree distribution in the real networks and their
$dK$-randomizations.} \label{DegreeDistribution}
\end{figure*}

\subsection{Metrics defined by $dK$-distributions}\label{sec:dk-metrics}

We first consider the most basic metrics, which are defined by the
appropriate $dK$-distributions. Therefore it is not surprising that
$dK$-random graphs with appropriate $d$ have the values of these
metrics equal exactly to those in the real networks. Nevertheless,
we report these results for consistency and illustration purposes.

\subsubsection{$1K$: degree distribution}

Fig.~\ref{DegreeDistribution} shows the distributions $P(k)$
of node degrees $k$:
\begin{equation}
P(k)=\frac{N(k)}{N},
\end{equation}
where $N(k)$ is the number of nodes of degree $k$ in the network,
and $N$ is the total number of nodes in it, so that $P(k)$ is
normalized, $\sum_kP(k)=1$ (we do not consider nodes of degree $k=0$).
The $1K$-distribution fully defines the
$0K$-distribution, i.e., the average degree $\bar{k}$ in the
network, by
\begin{equation}\label{eq:bar_k}
\bar{k}=\sum_kkP(k),
\end{equation}
but not {\it vice versa}.

We observe in Fig.~\ref{DegreeDistribution} that while
$0K$-randomizations are off, the $1K$-random graphs reproduce the
degree distributions in the real networks exactly, which is by
dentition: the $1K$-distribution is the degree distribution, and
$1K$-randomization does not alter it. The $dK$-randomizations with
$d>1$ do not alter the $1K$-distribution either, therefore they also
match the degree distributions in the real networks exactly (not
shown).

\subsubsection{$2K$: average neighbor degree}

\begin{figure*}
\centerline{\includegraphics[width=.9\linewidth]{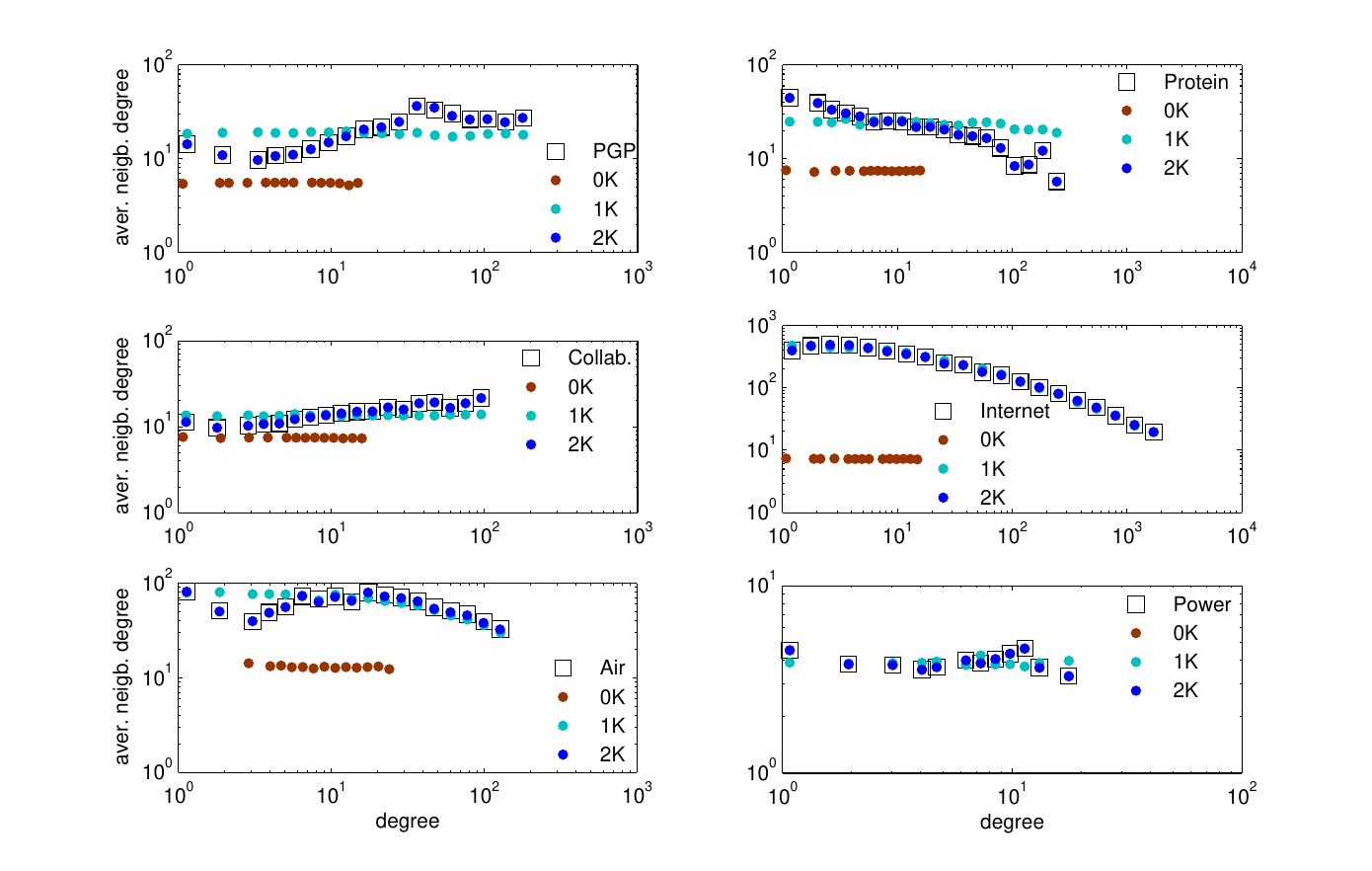}}
\caption{The average neighbour degree in the real networks and their
$dK$-randomizations.} \label{ANND}
\end{figure*}

\begin{figure*}
\centerline{\includegraphics[width=.9\linewidth]{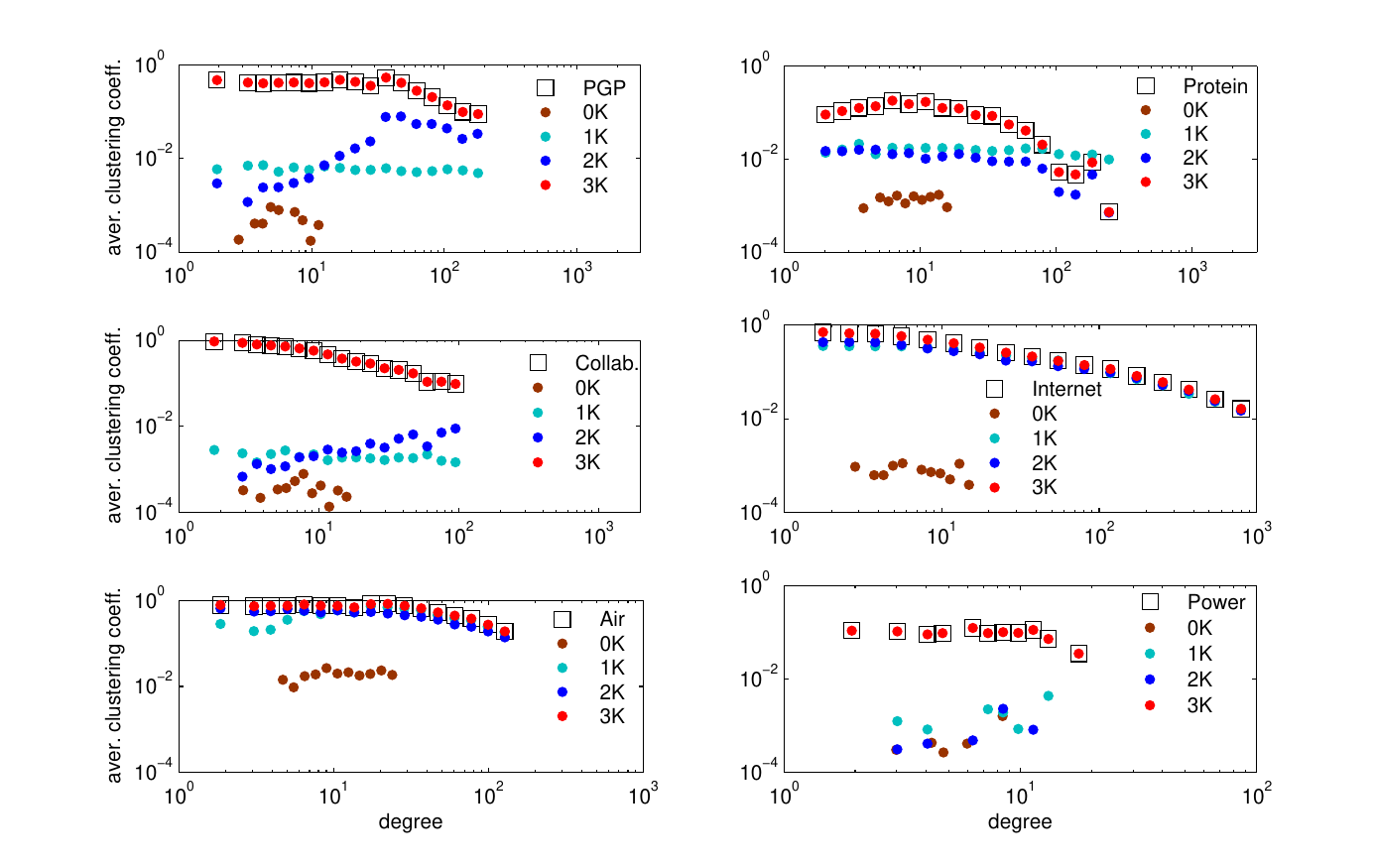}}
\caption{The degree-dependent clustering in the real networks and
their $dK$-randomizations.}
\label{DegreeDependendClusteringCoefficient}
\end{figure*}

Fig.~\ref{ANND} shows the average degree $\bar{k}_{nn}(k)$ of
neighbors of nodes of degree $k$. This function is a commonly used
projection of the joint degree distribution (JDD) $P(k,k')$, i.e.,
the $2K$-distribution. The JDD is defined as
\begin{equation}
P(k,k')=\mu(k,k')\frac{N(k,k')}{2M},
\end{equation}
where $N(k,k')=N(k',k)$ is the number of links between nodes of
degrees $k$ and $k'$ in the network, $M$ is the total number of
links in it, and
\begin{equation}
\mu(k,k')=
\begin{cases}
2&\text{if $k=k'$,}\\
1&\text{otherwise,}
\end{cases}
\end{equation}
so that $P(k,k')$ is normalized, $\sum_{k,k'}P(k,k')=1$. The
$2K$-distribution fully defines the $1K$-distribution by
\begin{equation}
P(k)=\frac{\bar{k}}{k}\sum_{k'}P(k,k'),
\end{equation}
but not {\it vice versa}. The average neighbor degree
$\bar{k}_{nn}(k)$ is a projection of the $2K$-distribution $P(k,k')$
via
\begin{equation}
\bar{k}_{nn}(k)=\frac{\bar{k}}{kP(k)}\sum_{k'}k'P(k,k')=
\frac{\sum_{k'}k'P(k,k')}{\sum_{k'}P(k,k')}.
\end{equation}

We observe in Fig.~\ref{ANND} that while $0K$-randomizations are way
off, the $1K$-randomization are much closer to the real networks,
whereas the $2K$-randomizations have exactly the same average
neighbor degrees as the real networks, which is again by definition:
$2K$-randomization does not change $P(k,k')$. In the Internet case,
even $1K$-randomization does not noticeably affect
$\bar{k}_{nn}(k)$. The $dK$-randomizations with $d>2$ do not alter
$P(k,k')$ and consequently $\bar{k}_{nn}(k)$ at all, therefore they
reproduce the latter exactly as well for all the networks (not
shown).

\subsubsection{$3K$: clustering}

Fig.~\ref{DegreeDependendClusteringCoefficient} shows
degree-dependent clustering $\bar{c}(k)$. Clustering of node $i$ is
the number of triangles $\triangle_i$ it forms, or equivalently the
number of links among its neighbors, divided by the maximum such
number, which is $k(k-1)/2$, where $k$ is $i$'s degree, $\deg(i)=k$.
Averaging over all nodes of degree $k$, the degree-dependent
clustering is
\begin{equation}\label{eq:clustering-definition}
\bar{c}(k) = \frac{2\triangle(k)}{k(k-1)N(k)},\;\text{where}\;
\triangle(k) = \sum_{i:\,\deg(i)=k}\triangle_i.
\end{equation}

The degree-dependent clustering is a commonly used projection of the
$3K$-distribution. (See~\cite{SeBo06a,SeBo06b} for an alternative
formalism involving three point correlations.)

The $3K$-distribution is actually two
distributions characterizing the concentrations of the two
non-isomorphic degree-labeled subgraphs of size~$3$, wedges and
triangles:\\
\centerline{\includegraphics[width=2in]{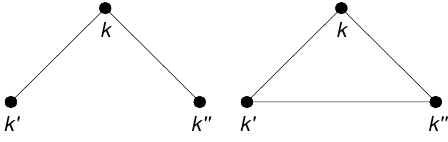}}. Let
$N_\wedge(k',k,k'')=N_\wedge(k'',k,k')$ be the number wedges
involving nodes of degrees $k$, $k'$, and $k''$, where $k$ is the
central node degree, and let $N_\triangle(k,k',k'')$ be the number
of triangles consisting of nodes of degrees $k$, $k'$, and $k''$,
where $N_\triangle(k,k',k'')$ is assumed to be symmetric with
respect to all permutations of its arguments. Then the two
components of the $3K$-distribution are
\begin{eqnarray}
P_\wedge(k',k,k'') &=& \mu(k',k'')\frac{N_\wedge(k',k,k'')}{2W},\\
P_\triangle(k,k',k'') &=& \nu(k,k',k'')\frac{N_\triangle(k,k',k'')}{6T},
\end{eqnarray}
where $T$ and $W$ are the total numbers of triangles and wedges in
the network, and
\begin{equation}
\nu(k,k',k'') =
\begin{cases}
6 & \text{if $k=k'=k''$},\\
1 & \text{if $k \neq k' \neq k''$},\\
2 & \text{otherwise},
\end{cases}
\end{equation}
so that both $P_\wedge(k',k,k'')$ and $P_\triangle(k,k',k'')$ are
normalized,
$\sum_{k,k',k''}P_\wedge(k',k,k'')=\sum_{k,k',k''}P_\triangle(k,k',k'')=1$.
The $3K$-distribution defines the $2K$-distribution (but not {\it
vice versa}), by
\begin{eqnarray}
P(k,k') &=& \frac{1}{k+k'-2}\sum_{k''}\left\{ \frac{6T}{M}P_\triangle(k,k',k'') \right.\nonumber\\
&+& \left. \frac{W}{M}\left[P_\wedge(k',k,k'')+P_\wedge(k,k',k'')\right] \right\}.
\end{eqnarray}
The normalization of $2K$- and $3K$-distributions implies the
following identity between the numbers of triangles, wedges, edges,
nodes, and the second moment of the degree distribution
$\bar{k^2}=\sum_k k^2 P(k)$:
\begin{equation}
2\frac{3T+W+M}{N}=\bar{k^2}.
\end{equation}
The degree-dependent clustering coefficient $\bar{c}(k)$ is the
following projection of the $3K$-distribution
\begin{equation}
\bar{c}(k) = \frac{6T}{N}\frac{\sum_{k',k''}P_\triangle(k,k',k'')}{k(k-1)P(k)}.
\end{equation}

We observe in Fig.~\ref{DegreeDependendClusteringCoefficient} that
clustering in the real networks and their $dK$-randomizations with
$d=3$ is exactly the same, which is again by definition. For $d<3$,
clustering differs drastically in many cases, except for the air
transportation network and especially the Internet. Therefore we can
say that the Internet is very close to being $1K$-random, i.e.,
fully defined by its degree distribution, as far as the $dK$-based
metrics considered in this section are concerned. Neither $3K$-, $2K$-, nor even
$1K$-randomization alter its $dK$-based (projection) metrics
noticeably.

\subsection{Motifs and their Z-scores}

\begin{figure*}
\subfigure{\includegraphics[width=2.3in]{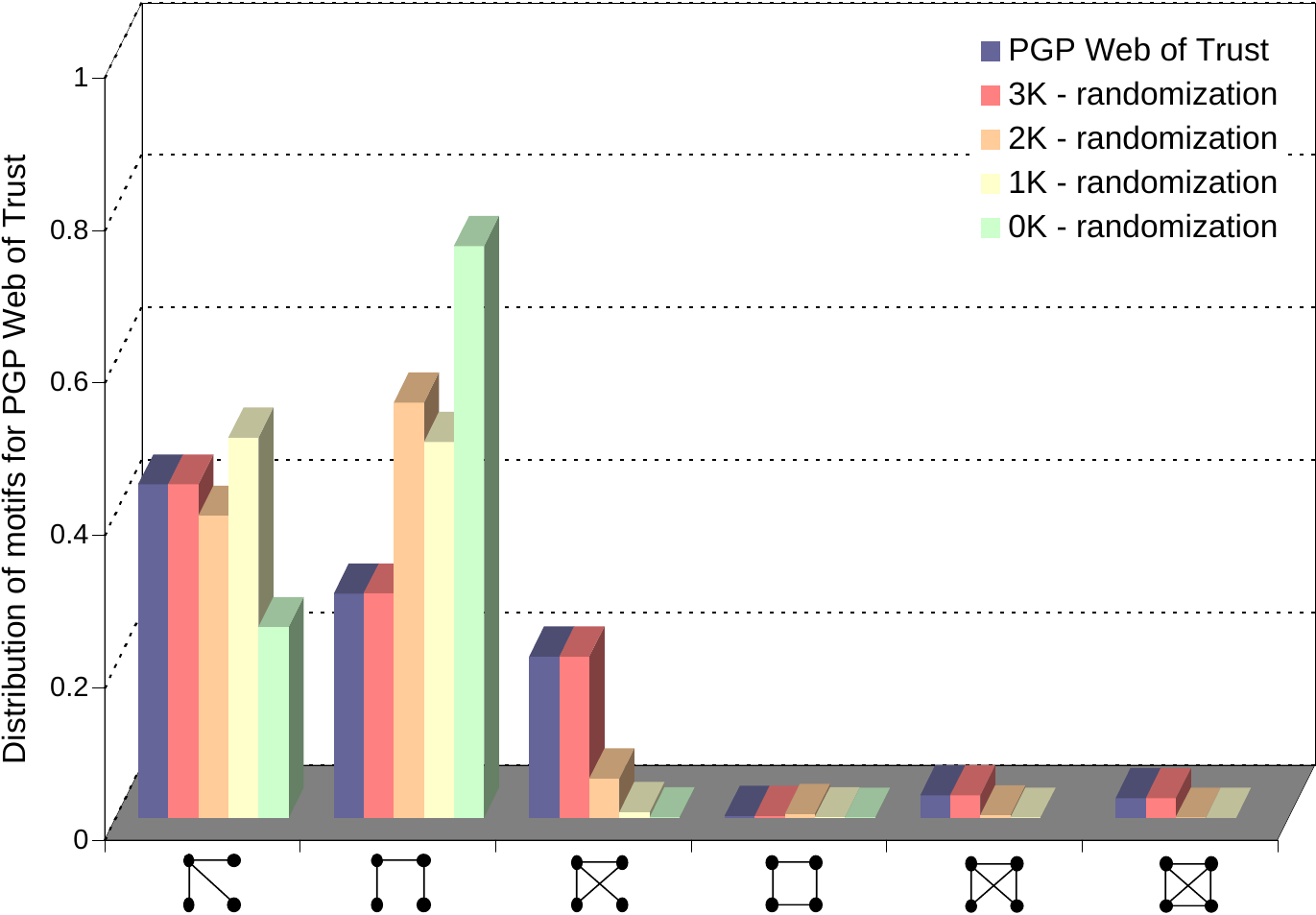}}
\subfigure{\includegraphics[width=2.3in]{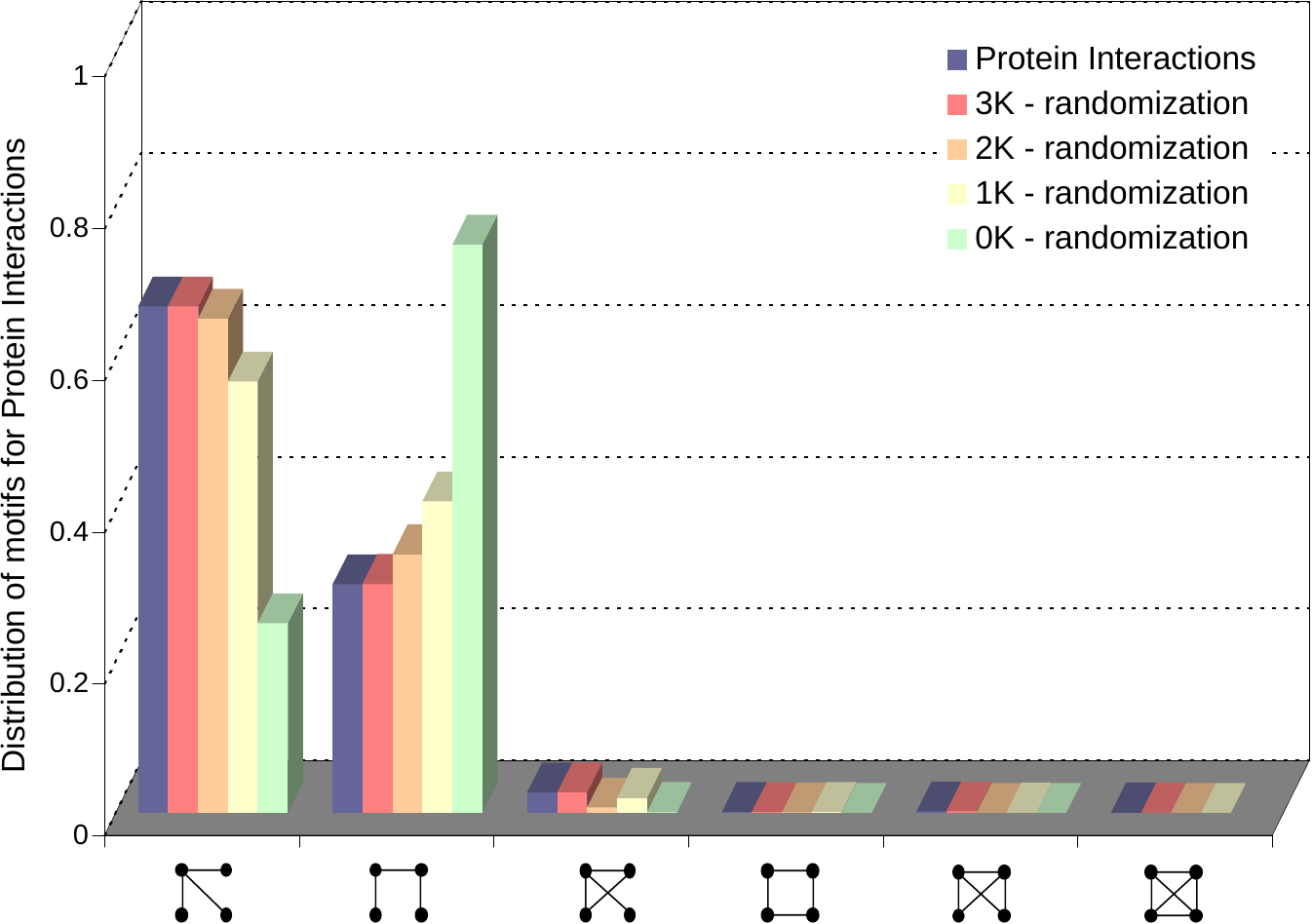}}
\subfigure{\includegraphics[width=2.3in]{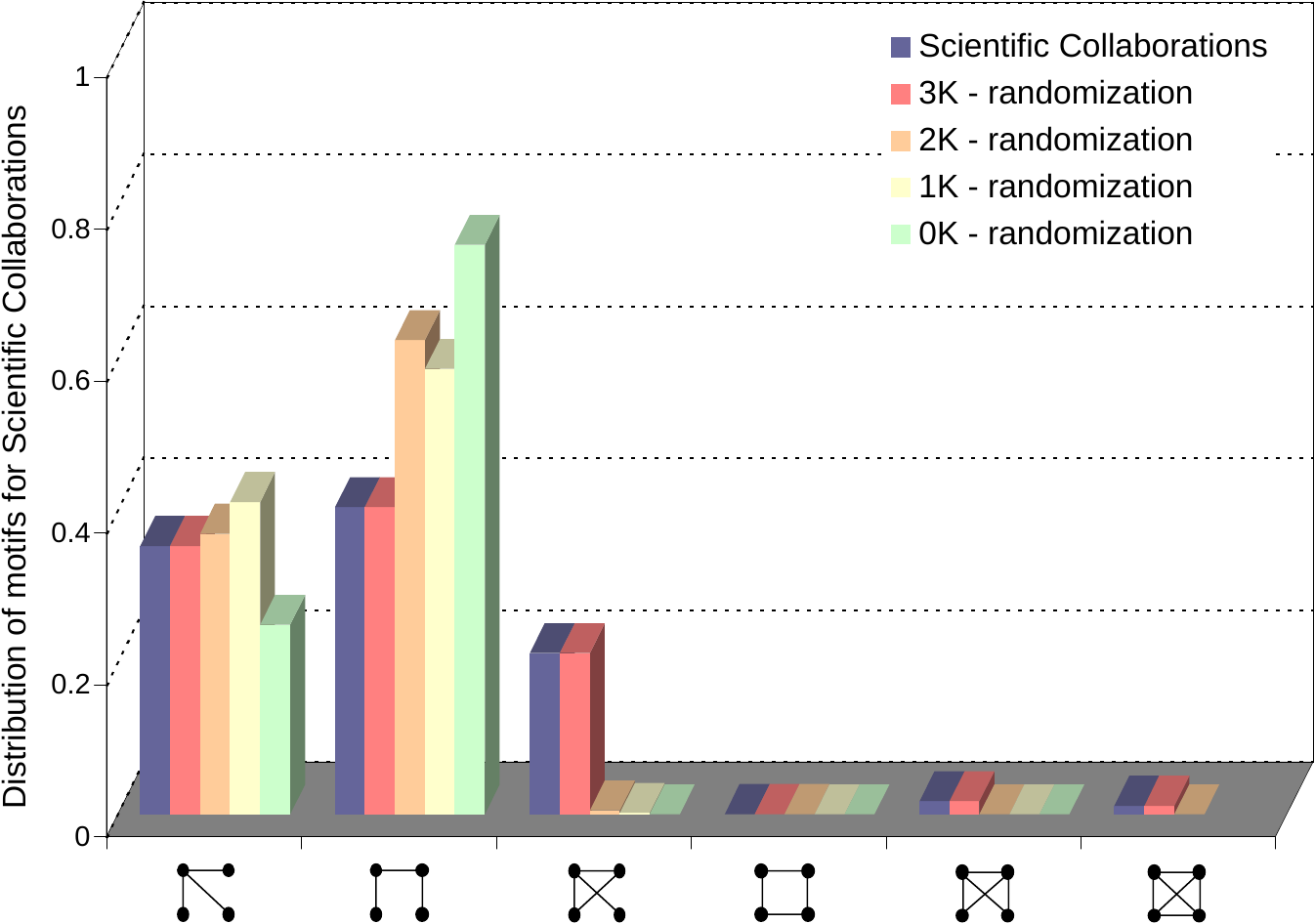}}\\
\subfigure{\includegraphics[width=2.3in]{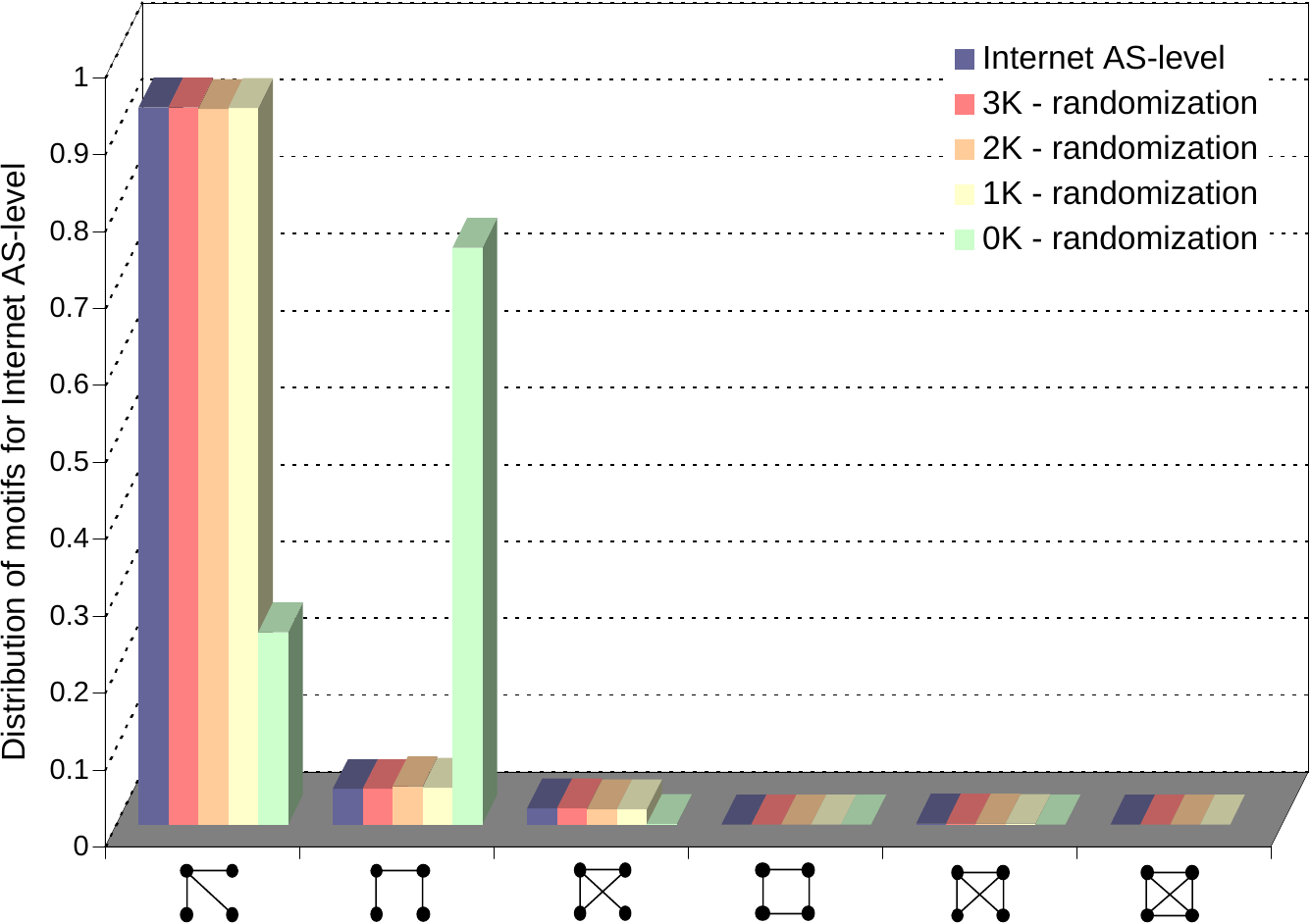}}
\subfigure{\includegraphics[width=2.3in]{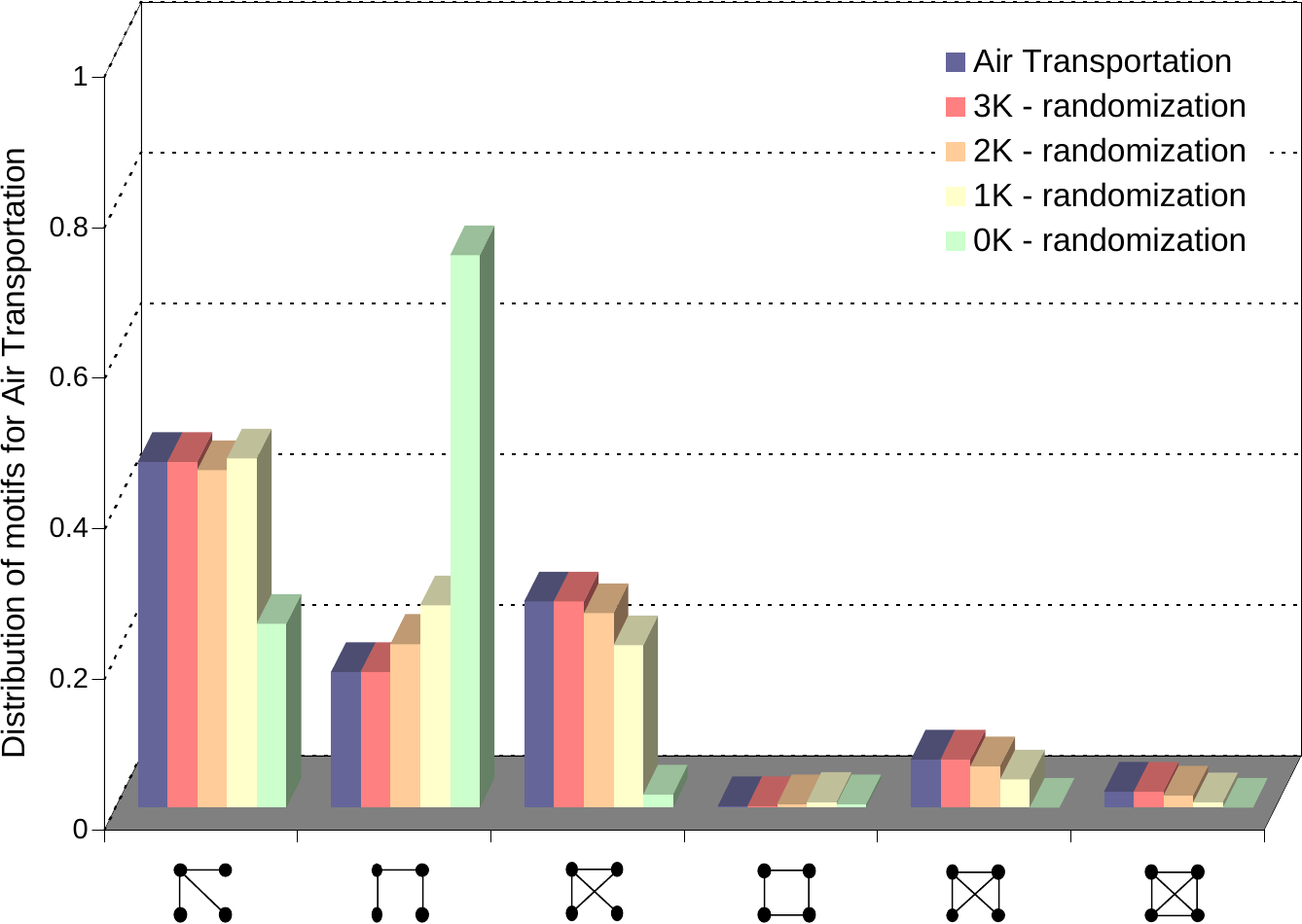}}
\subfigure{\includegraphics[width=2.3in]{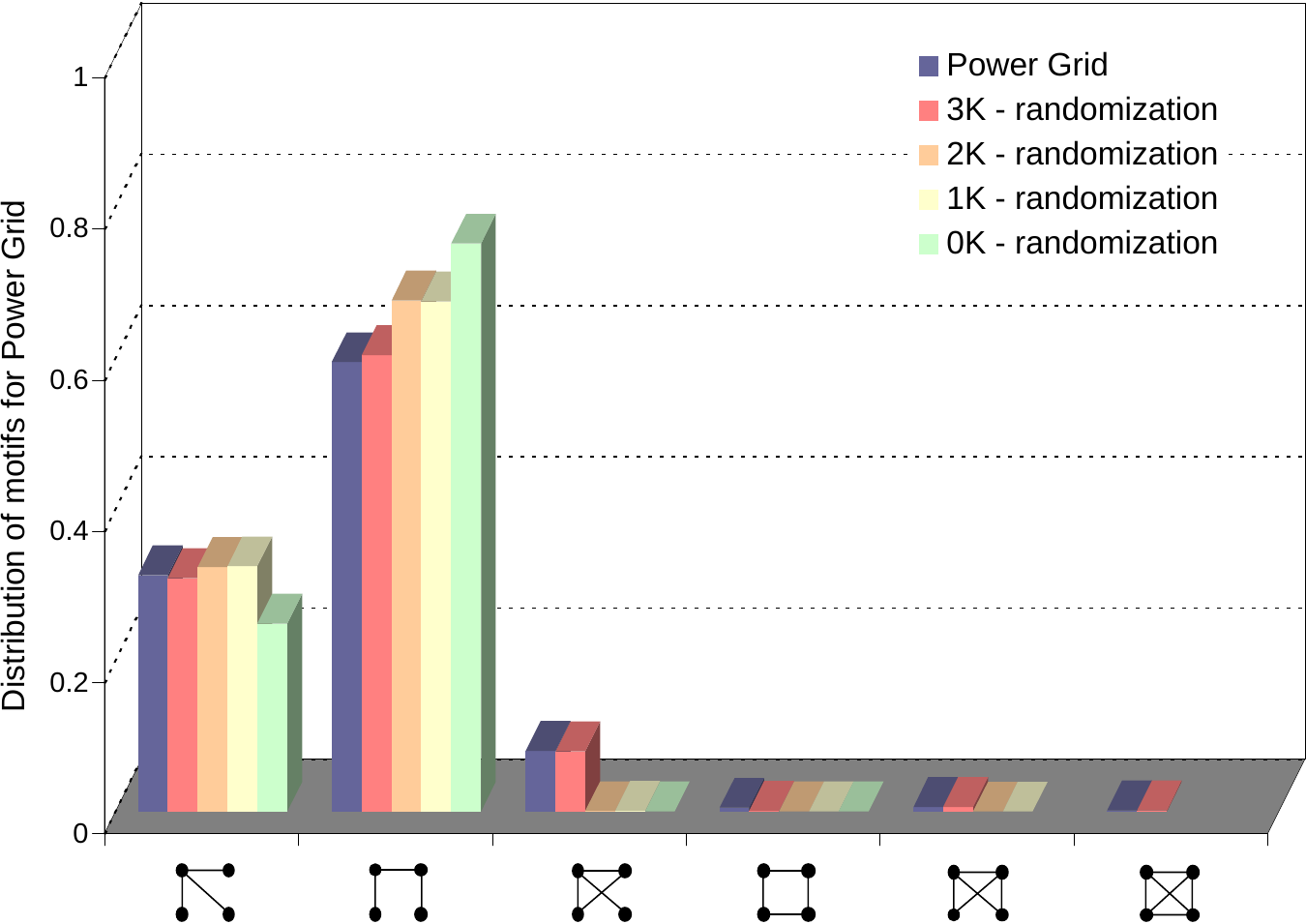}}
\caption{The motif distributions in the real networks and their
$dK$-randomizations.} \label{fig:motifs}
\end{figure*}
\begin{figure*}
\subfigure{\includegraphics[width=2.3in]{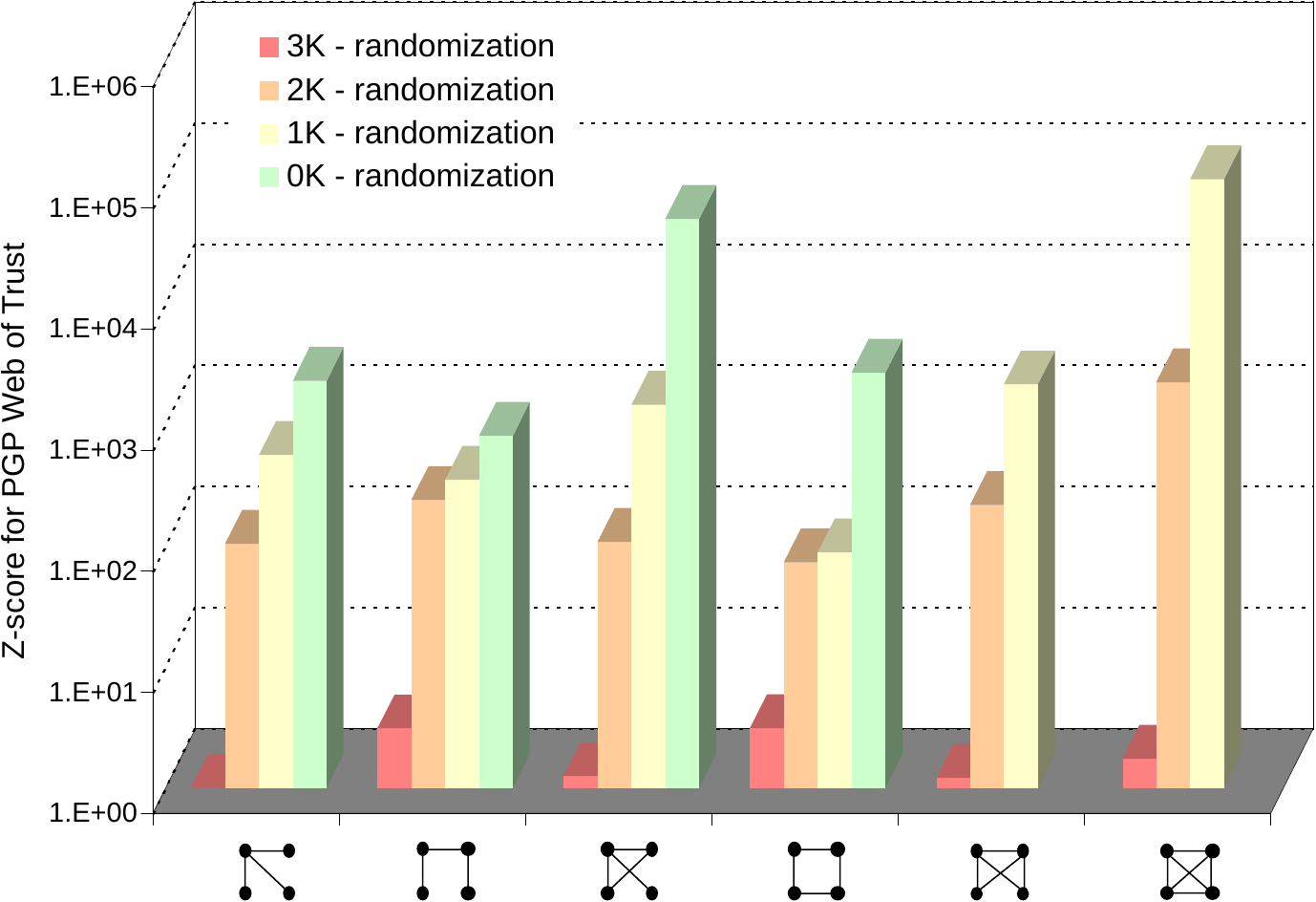}}
\subfigure{\includegraphics[width=2.3in]{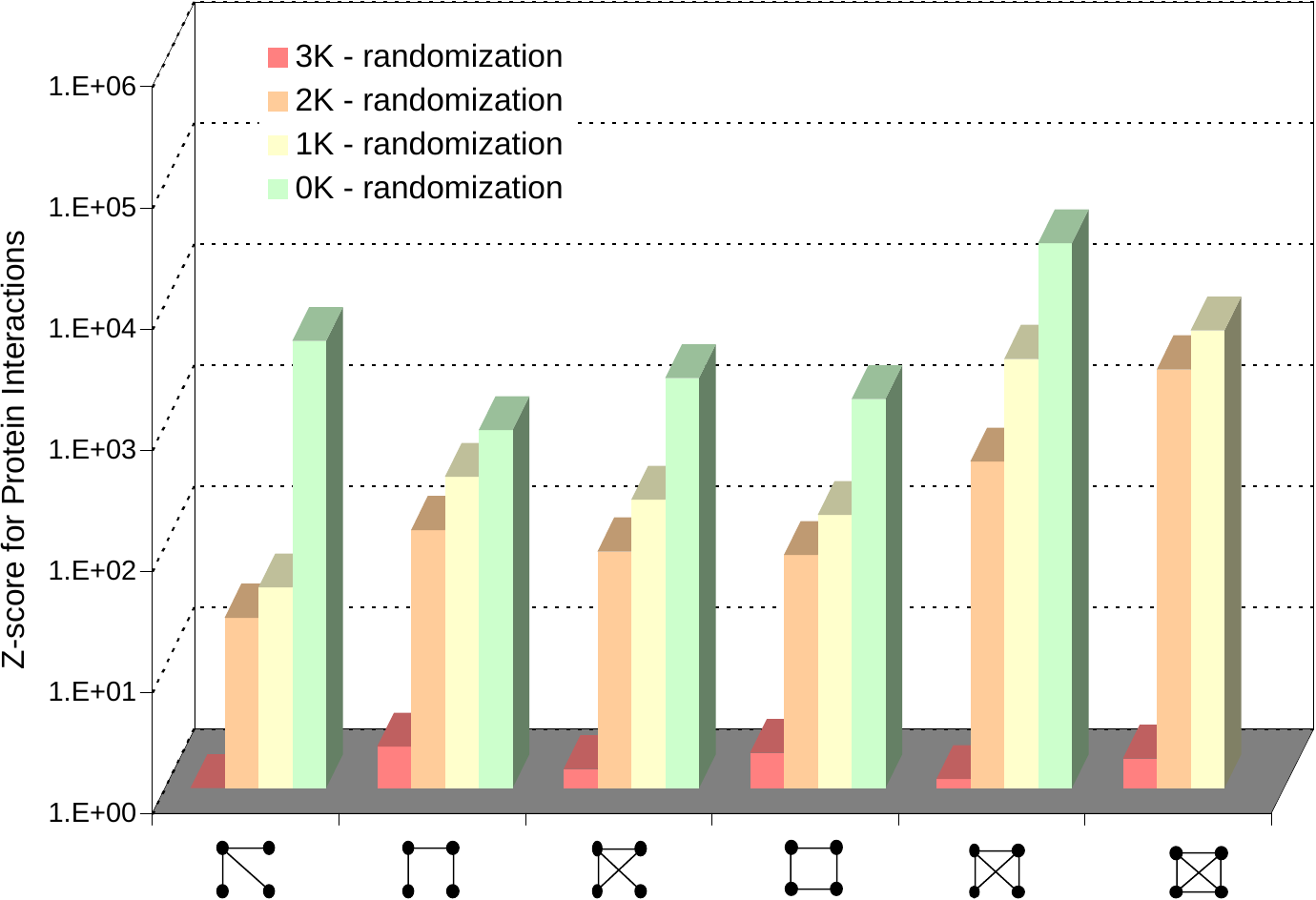}}
\subfigure{\includegraphics[width=2.3in]{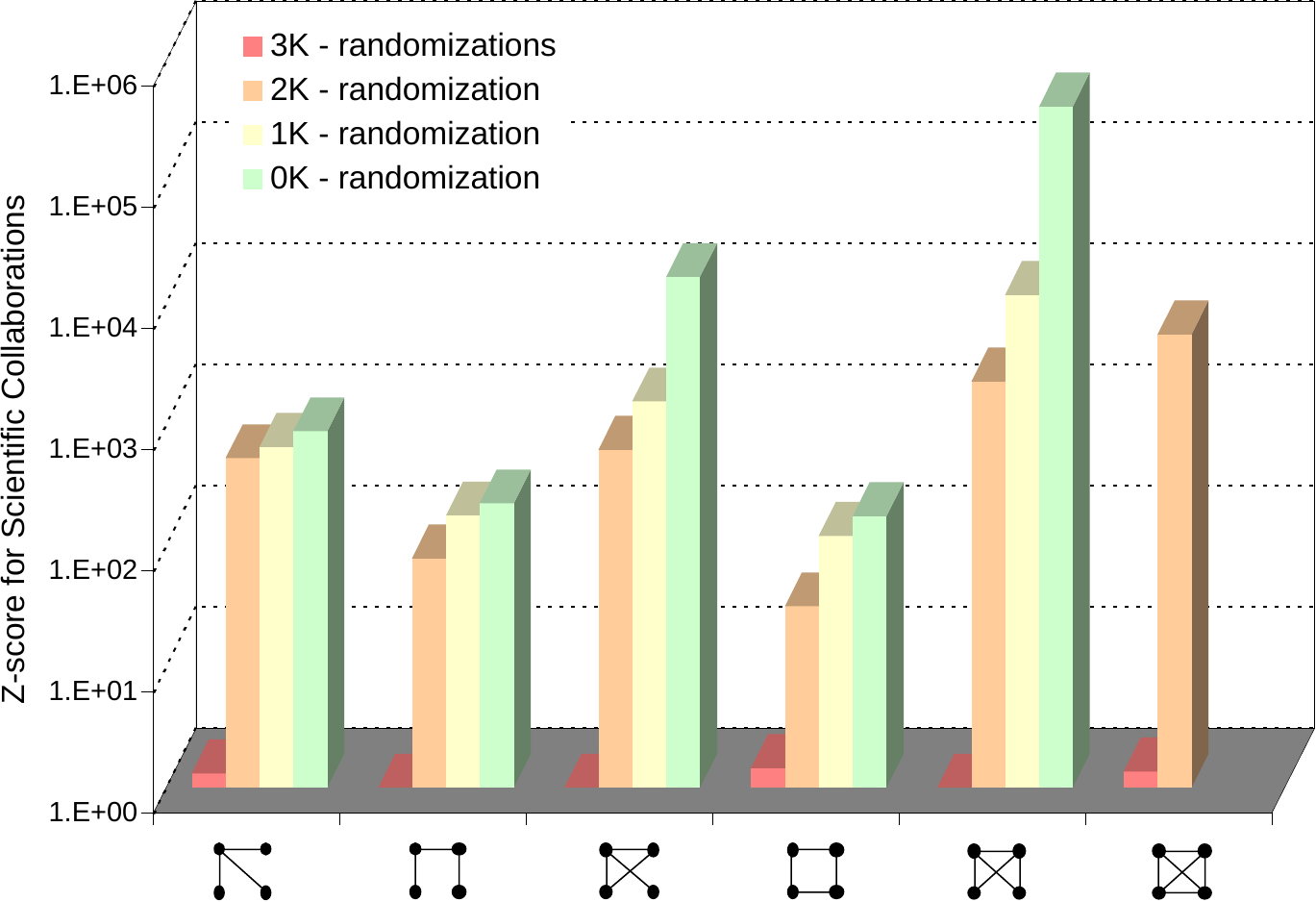}}\\
\subfigure{\includegraphics[width=2.3in]{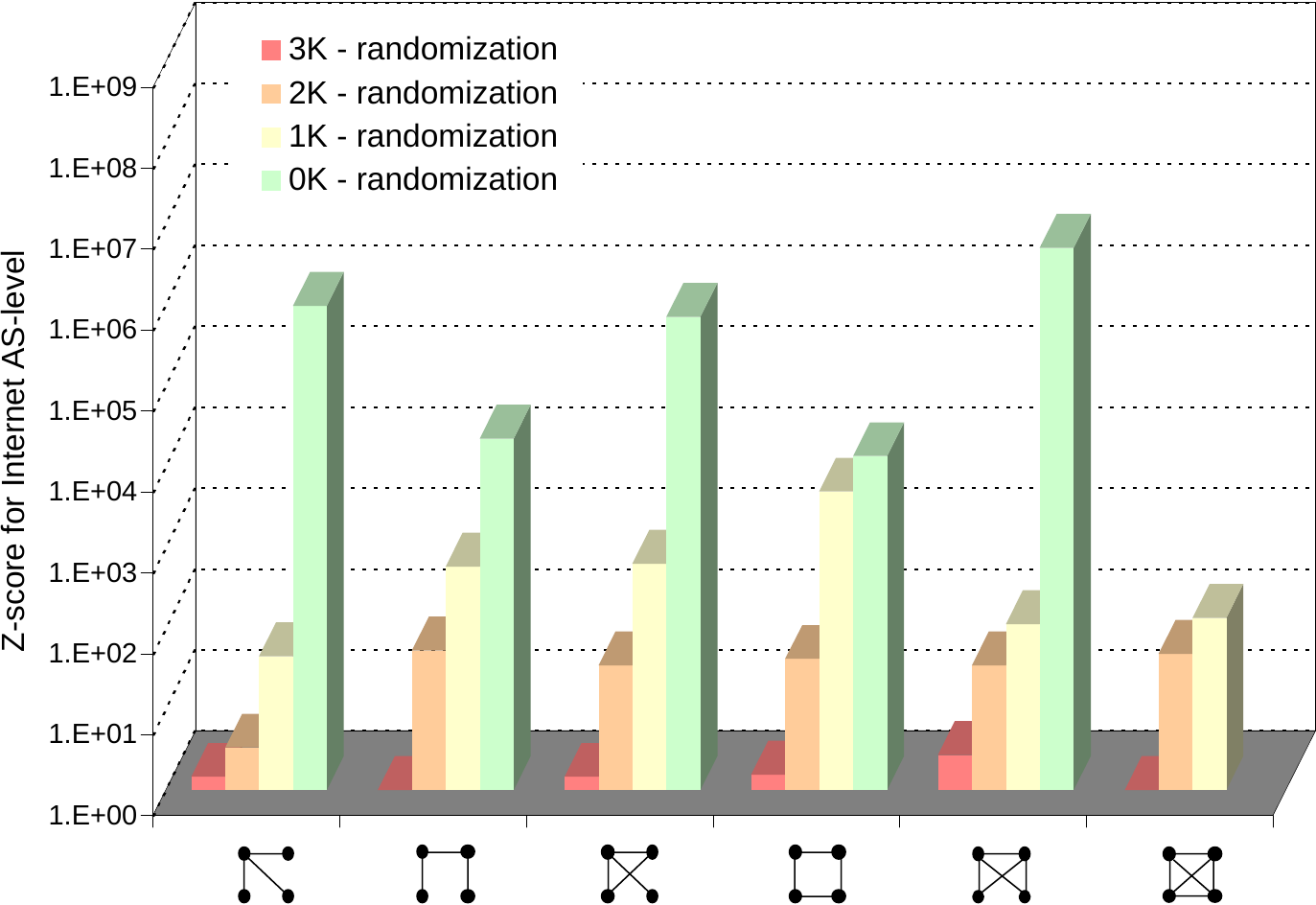}}
\subfigure{\includegraphics[width=2.3in]{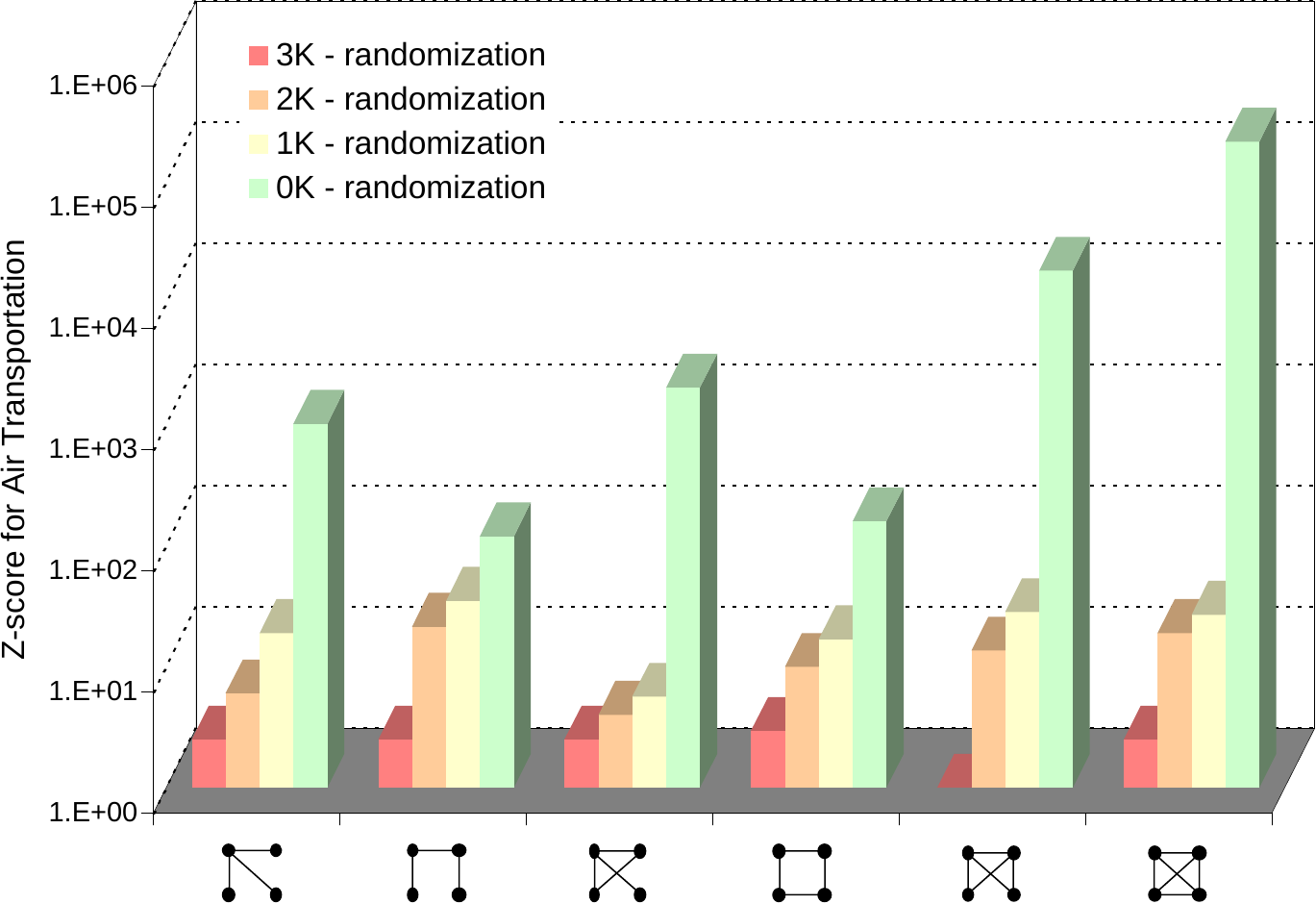}}
\subfigure{\includegraphics[width=2.3in]{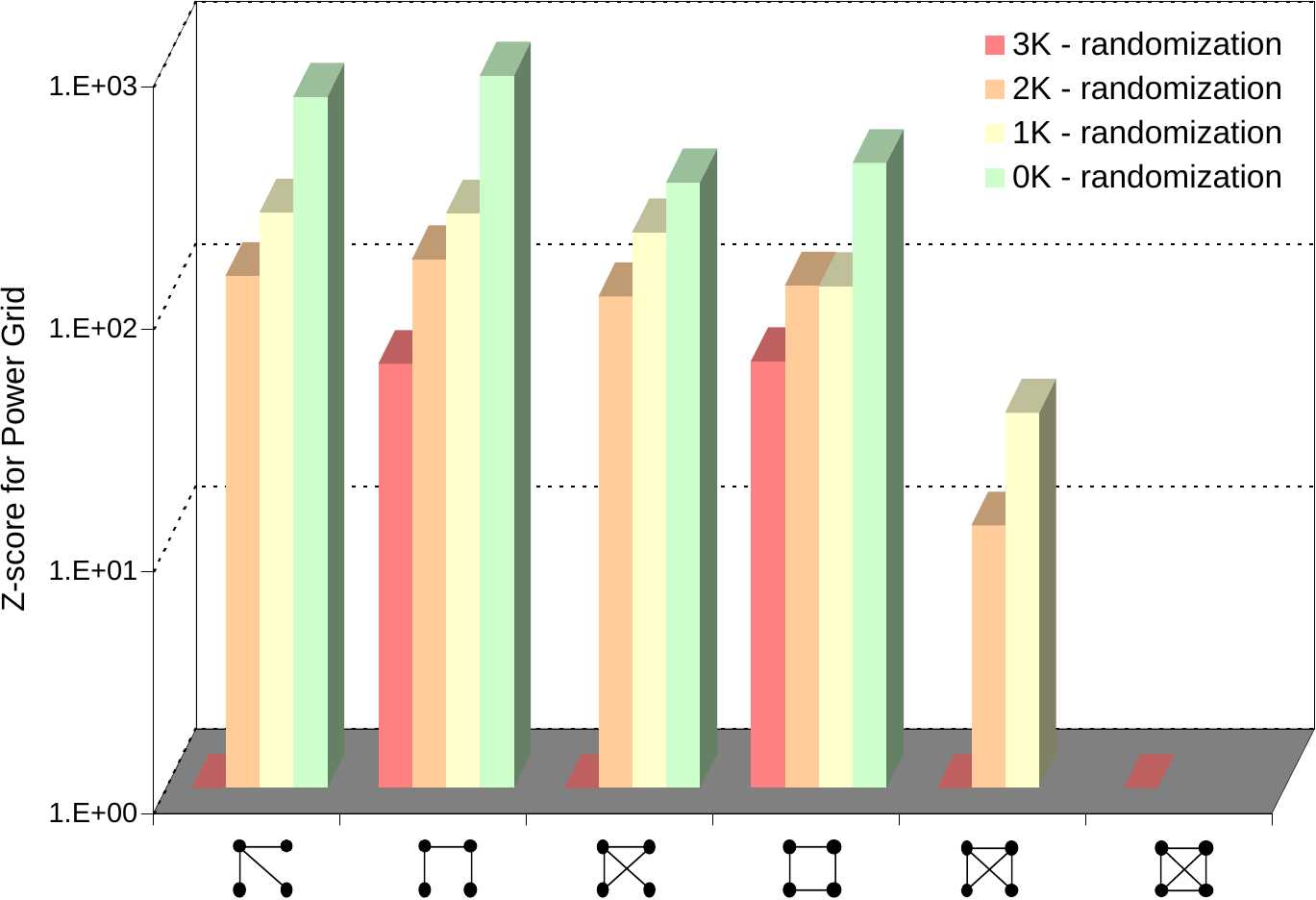}}
\caption{The motif Z-scores in the real networks and their
$dK$-randomizations.} \label{fig:z-scores}
\end{figure*}

There are six non-isomorphic motifs of size $4$, shown as the
$x$-axes in Figs.~\ref{fig:motifs},\ref{fig:z-scores}. For each
network and for each $d=0,1,2,3$, we obtain several $dK$-randomized
samples of the network, and then for each motif we compute its
distribution (normalized to the total number of subgraphs of size
$4$) in the real network, and its average distribution in the
$dK$-randomized samples of the network. The results are in
Fig.~\ref{fig:motifs}. Fig.~\ref{fig:z-scores} reports the
corresponding Z-scores. In certain cases, often for
$0K$-randomizations, some motifs do not occur at all in any
randomized samples, which explains the absence of some bars in the
figures.

The key observation is that when the randomization null model is
$3K$, the distributions of all motifs in the randomizations of all
the networks except the power grid, are close to those in the real
networks. The corresponding Z-scores are either low or zero. In
other words, all motifs are statistically non-significant.

\subsection{Distance and betweenness distributions}

\begin{figure*}
\centerline{\includegraphics[width=.75\linewidth]{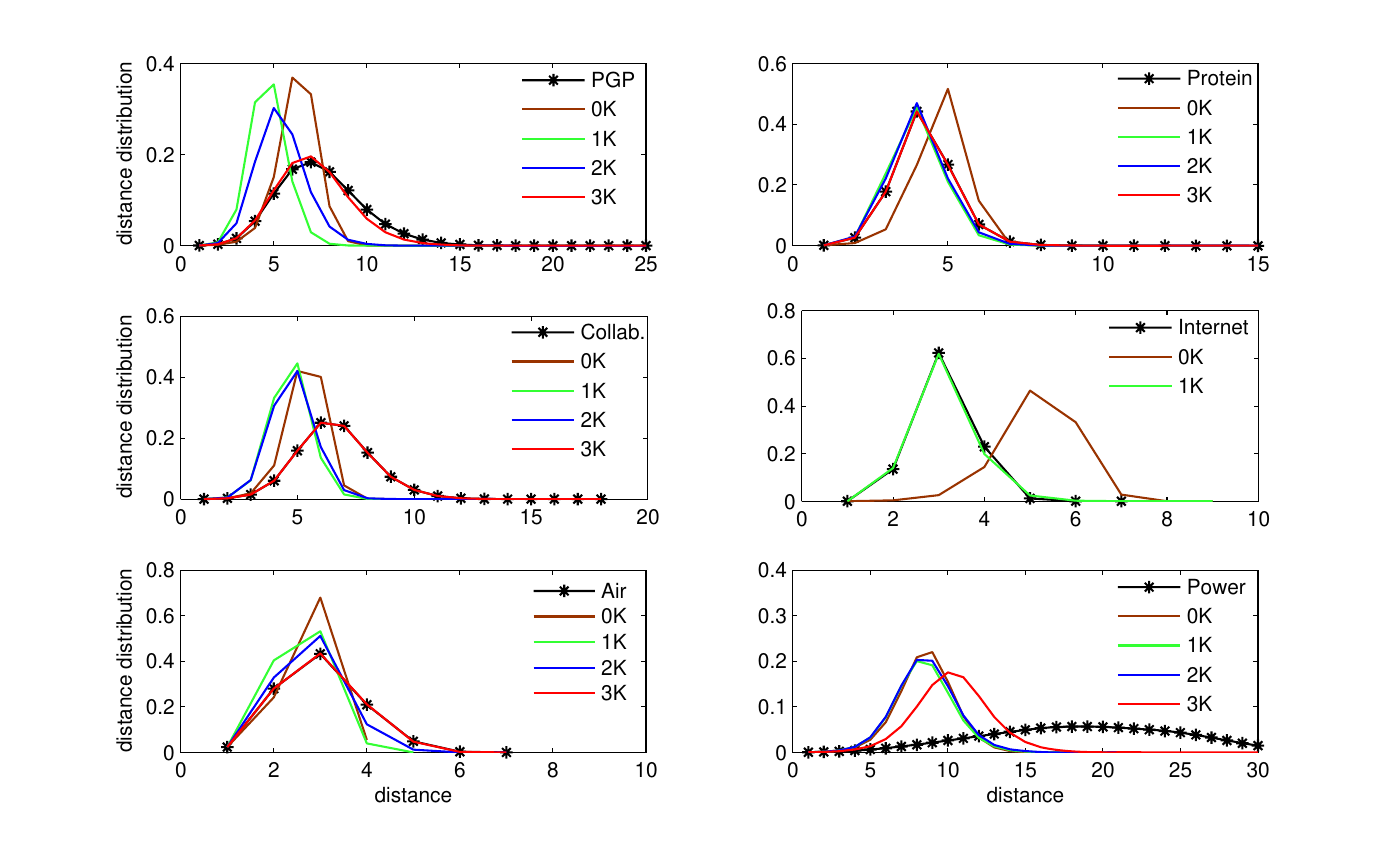}}
\caption{The distance distribution in the real networks and their
$dK$-randomizations.} \label{DistanceDistribution}
\end{figure*}
\begin{figure*}
\subfigure{\includegraphics[width=2.25in]{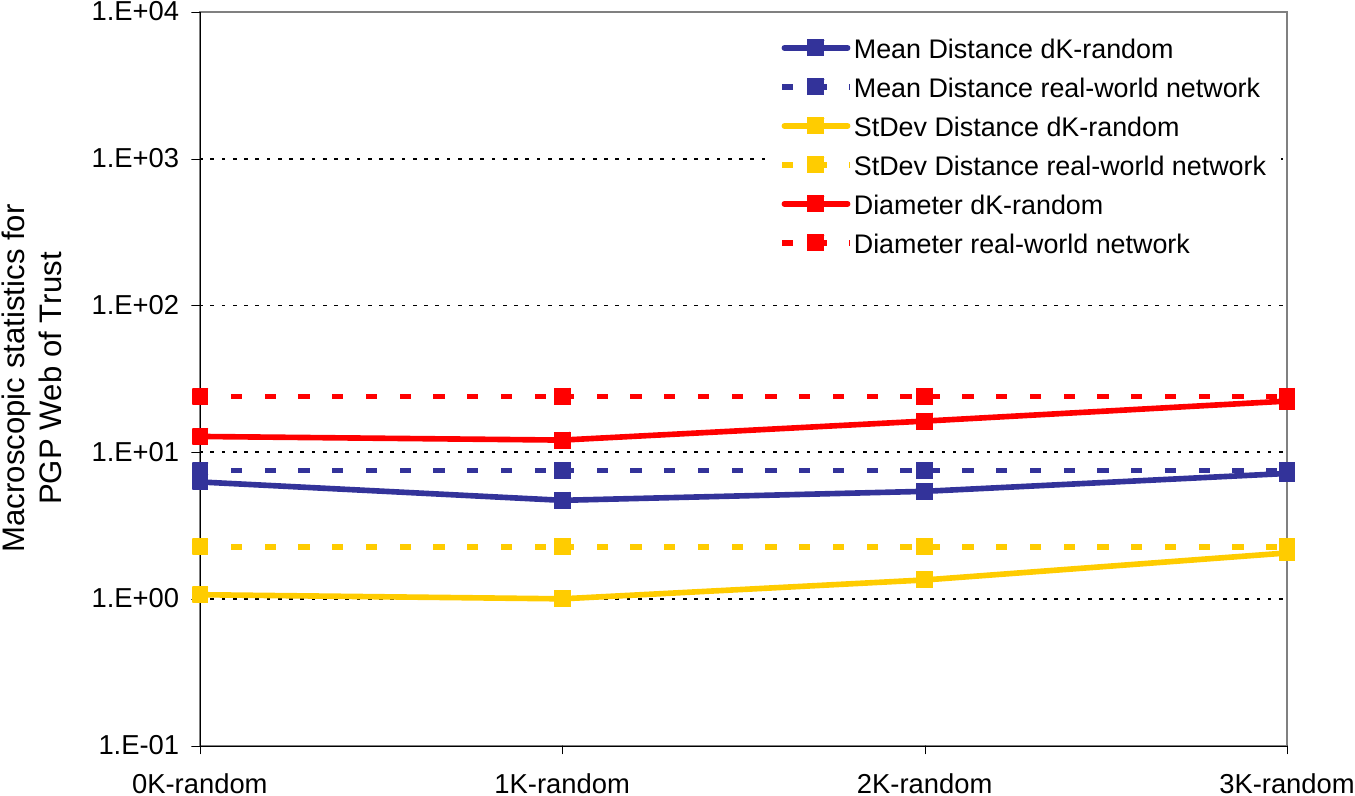}}
\subfigure{\includegraphics[width=2.25in]{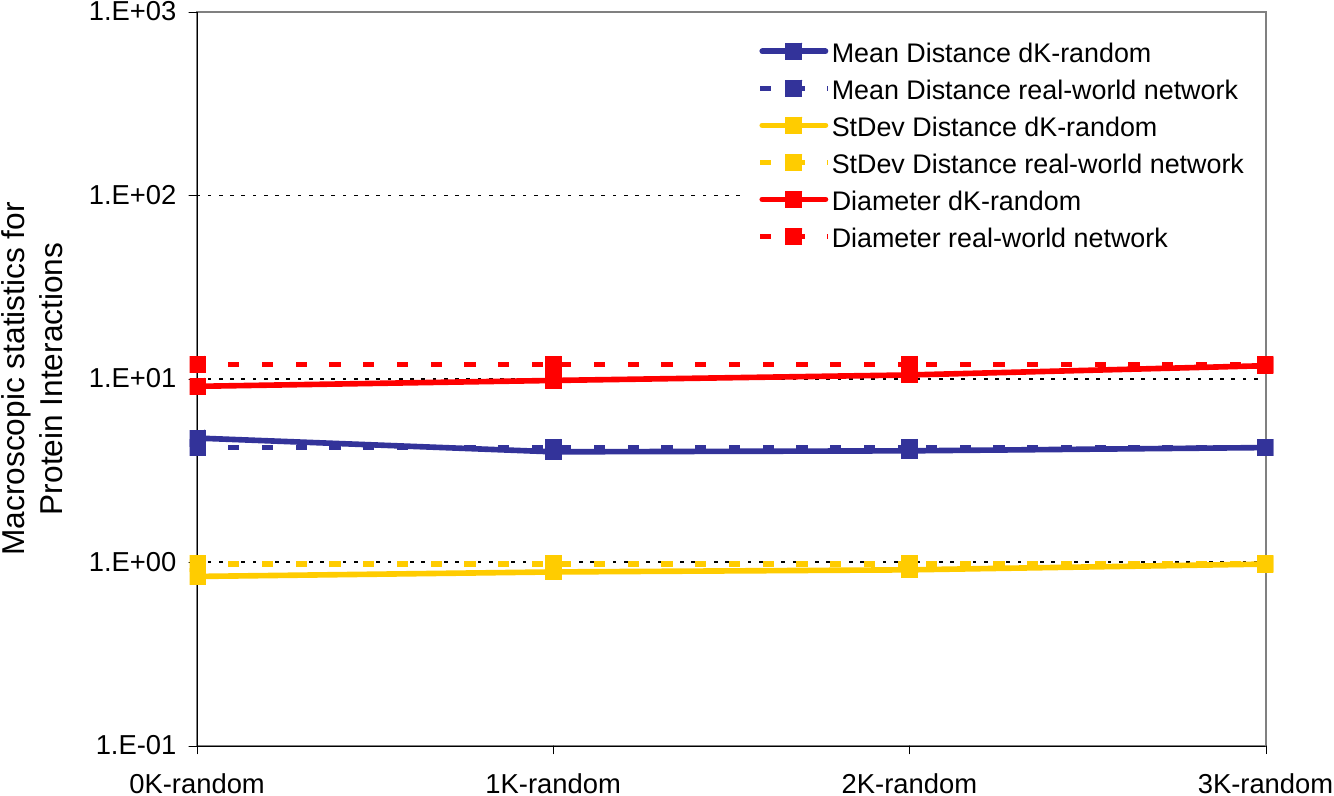}}\\
\subfigure{\includegraphics[width=2.25in]{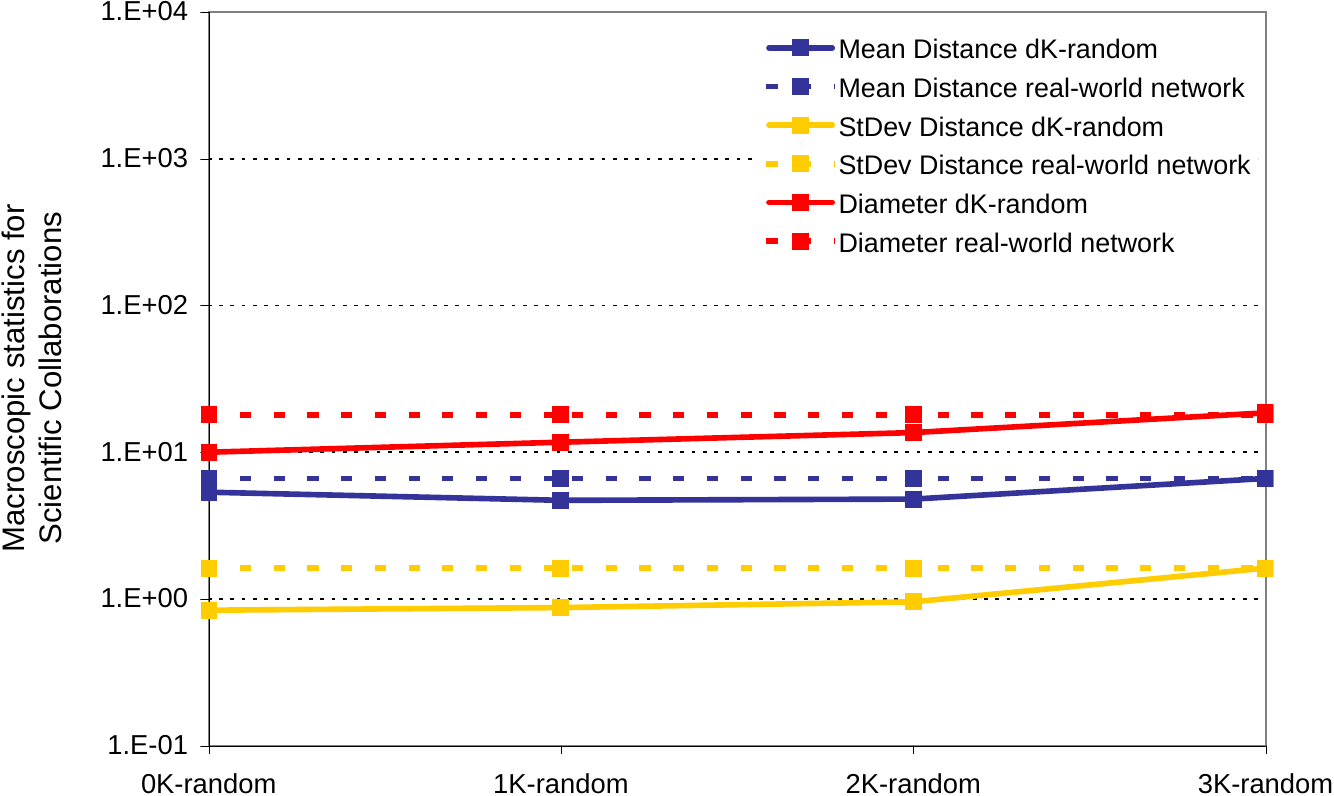}}
\subfigure{\includegraphics[width=2.25in]{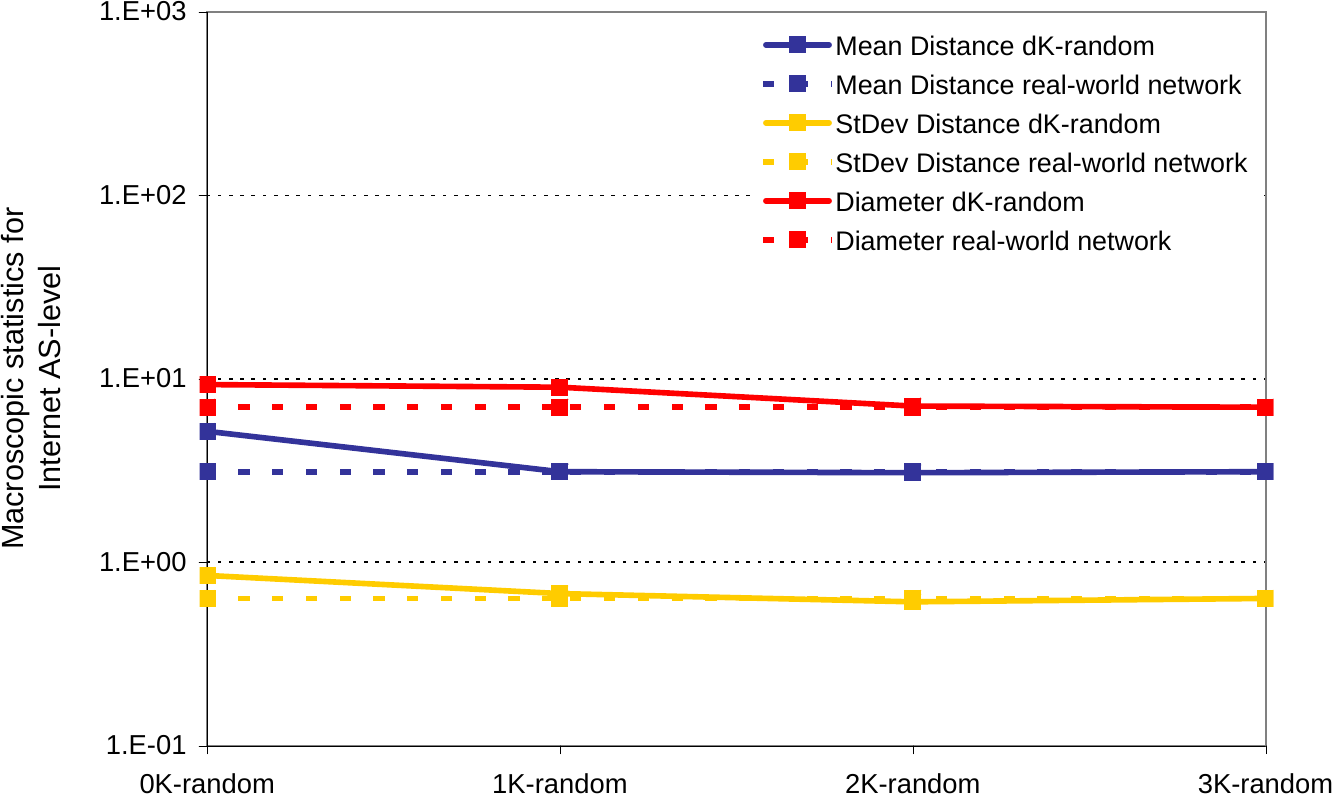}}\\
\subfigure{\includegraphics[width=2.25in]{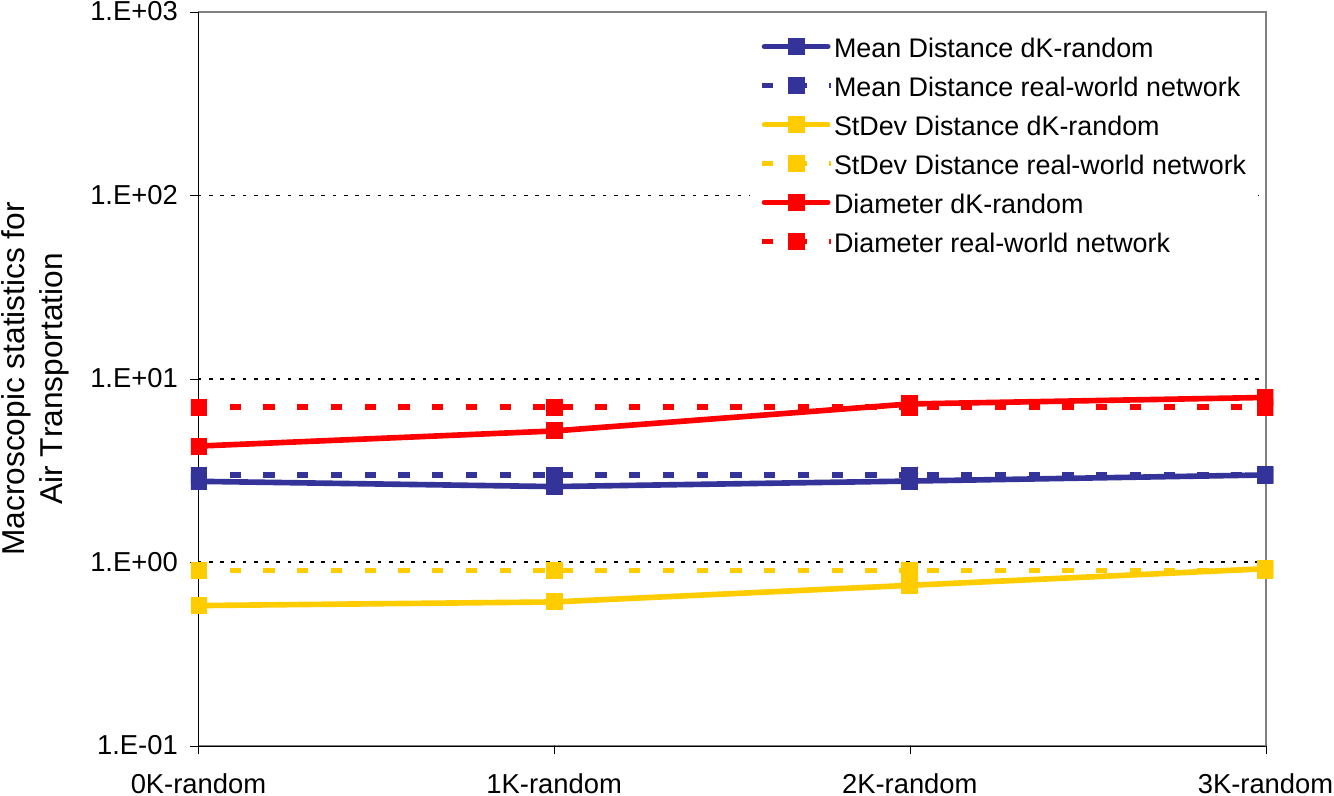}}
\subfigure{\includegraphics[width=2.25in]{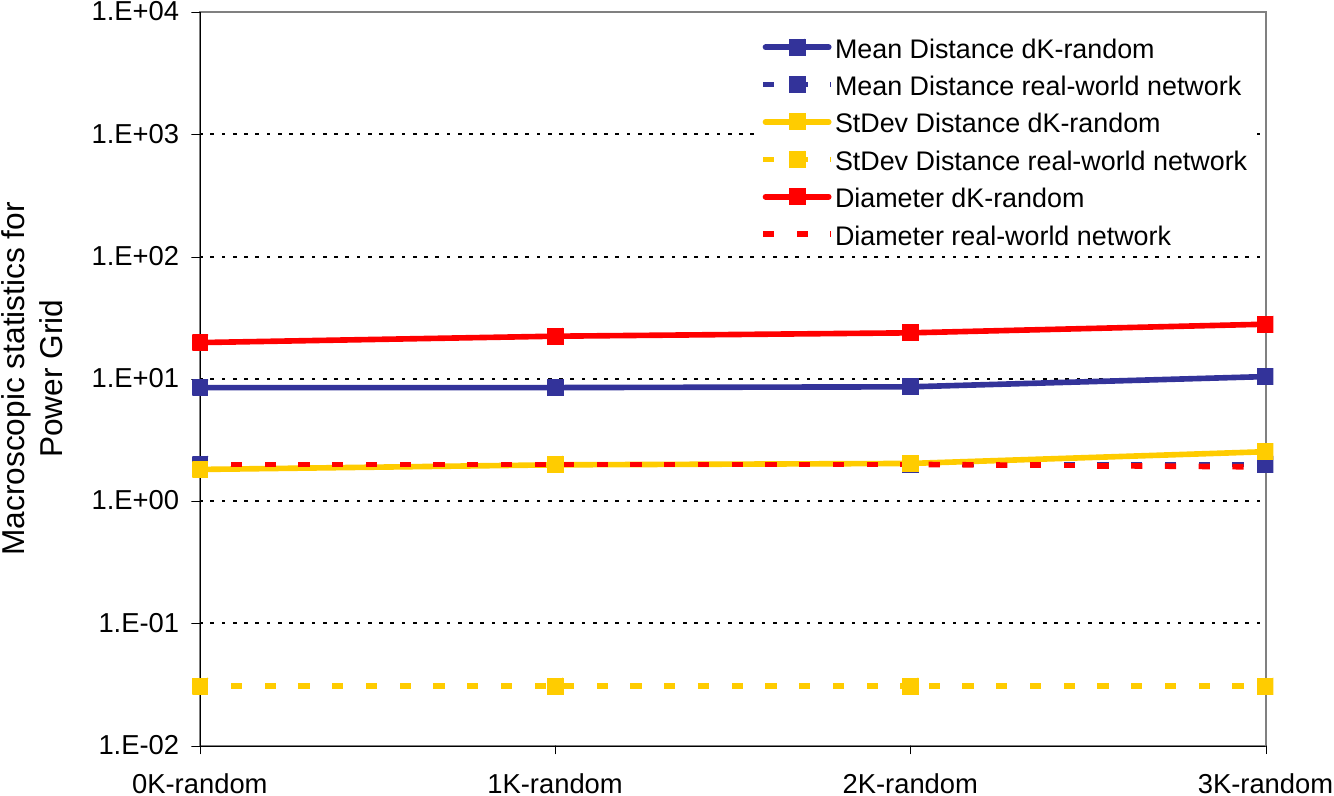}}\\
\caption{The average distance, the standard deviation of the
distance distribution, and the network diameter as functions of $d$
for $dK$-randomisations of the real networks. The corresponding
values for the real networks are shown by dashed lines.}
\label{DistanceDistribution(d)}
\end{figure*}

\begin{figure*}
\centerline{\includegraphics[width=.66\linewidth]{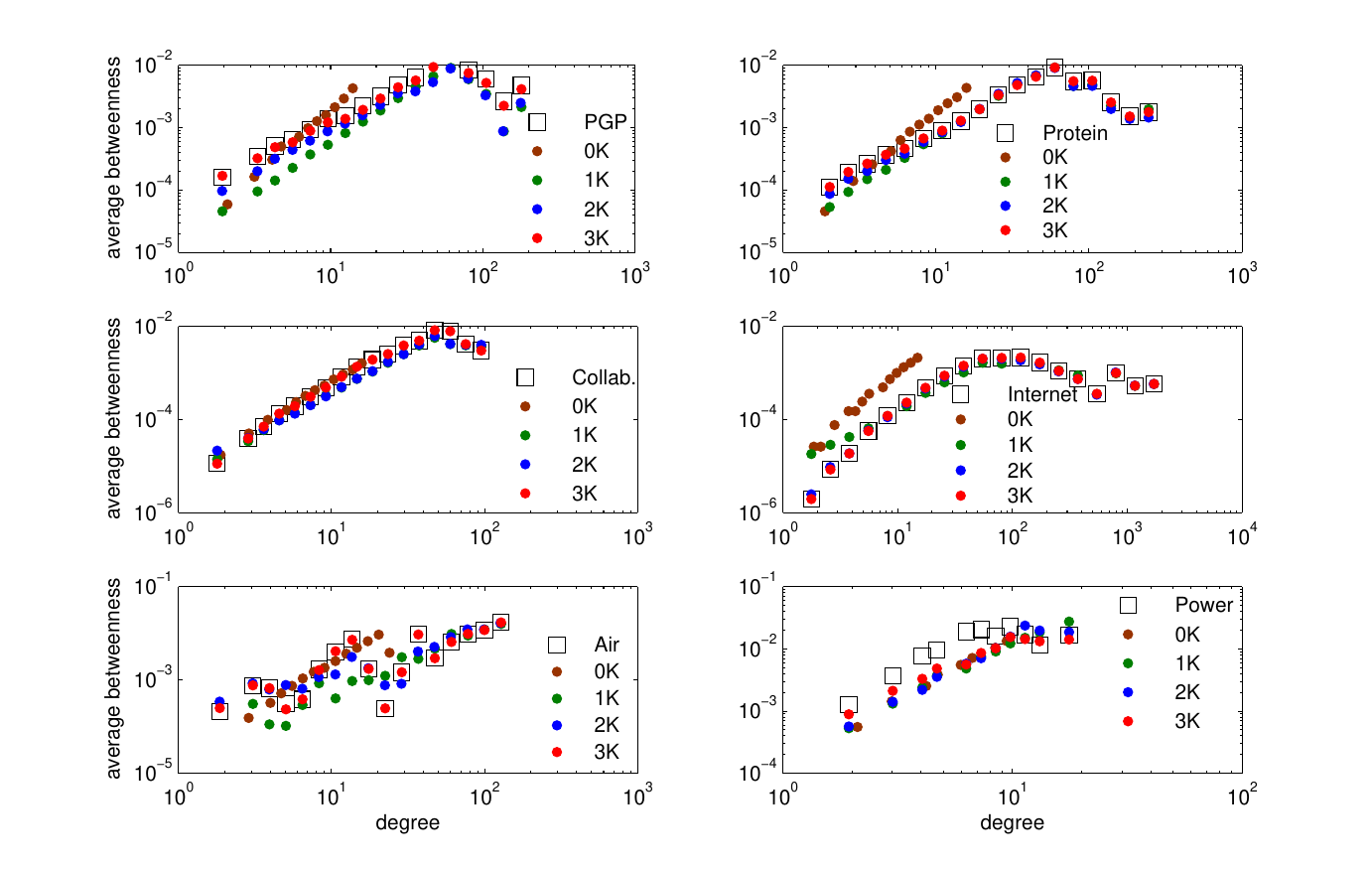}}
\caption{The average betweenness of nodes of a given degree in the
real networks and their $dK$-randomizations.}
\label{AverageBetweennessGivenDegree}
\end{figure*}
\begin{figure*}
\subfigure{\includegraphics[width=2.25in]{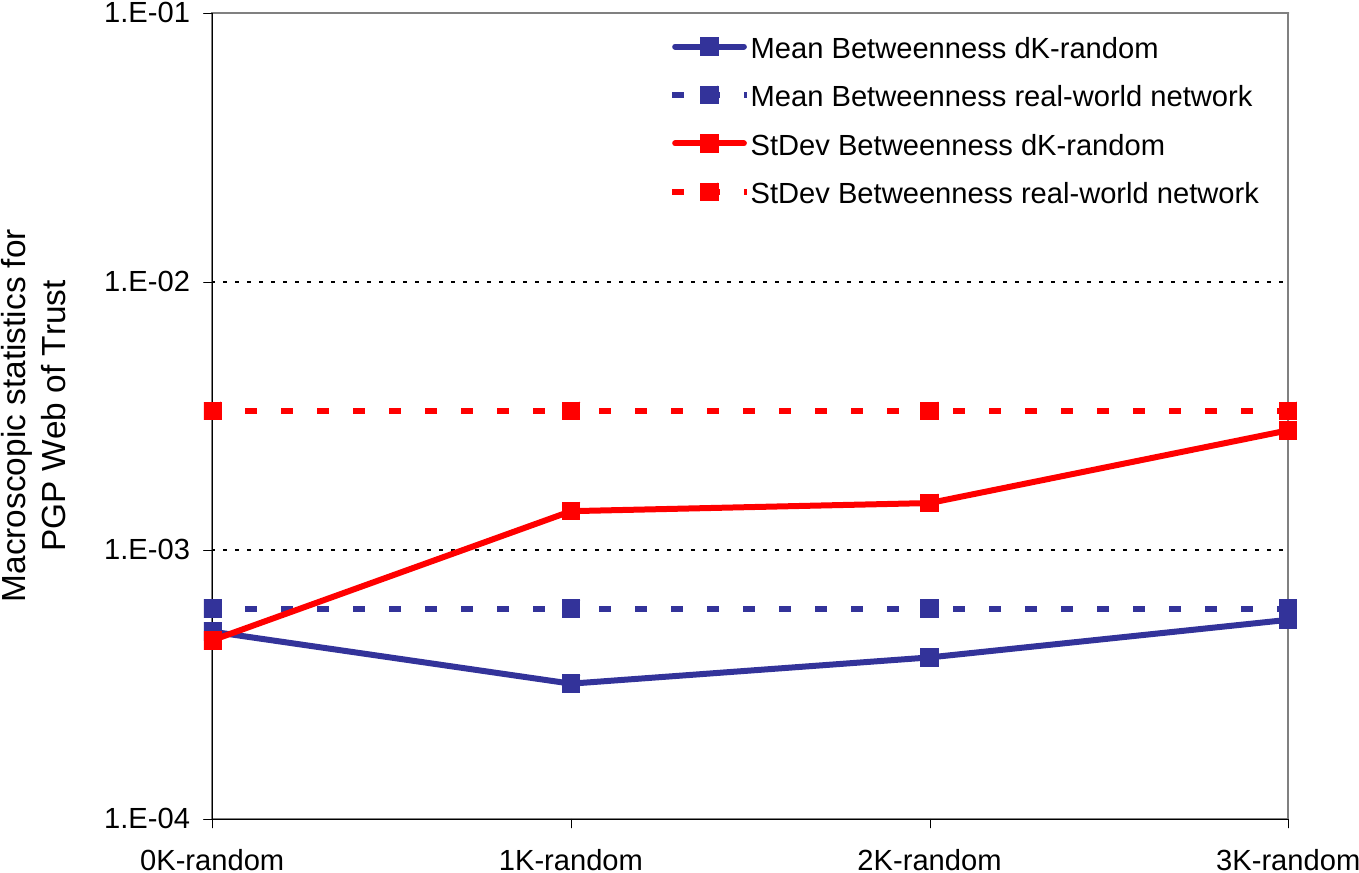}}
\subfigure{\includegraphics[width=2.25in]{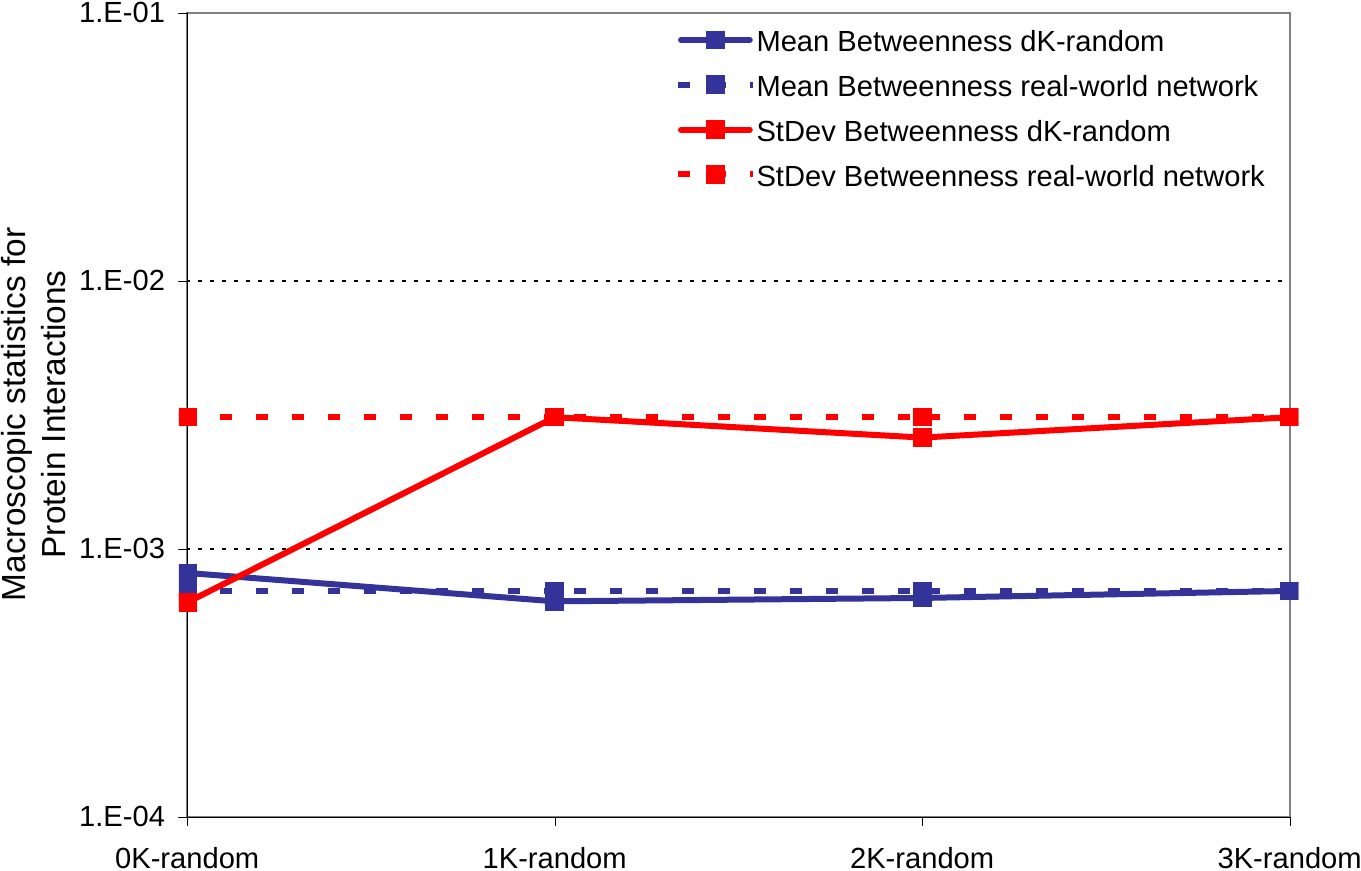}}\\
\subfigure{\includegraphics[width=2.25in]{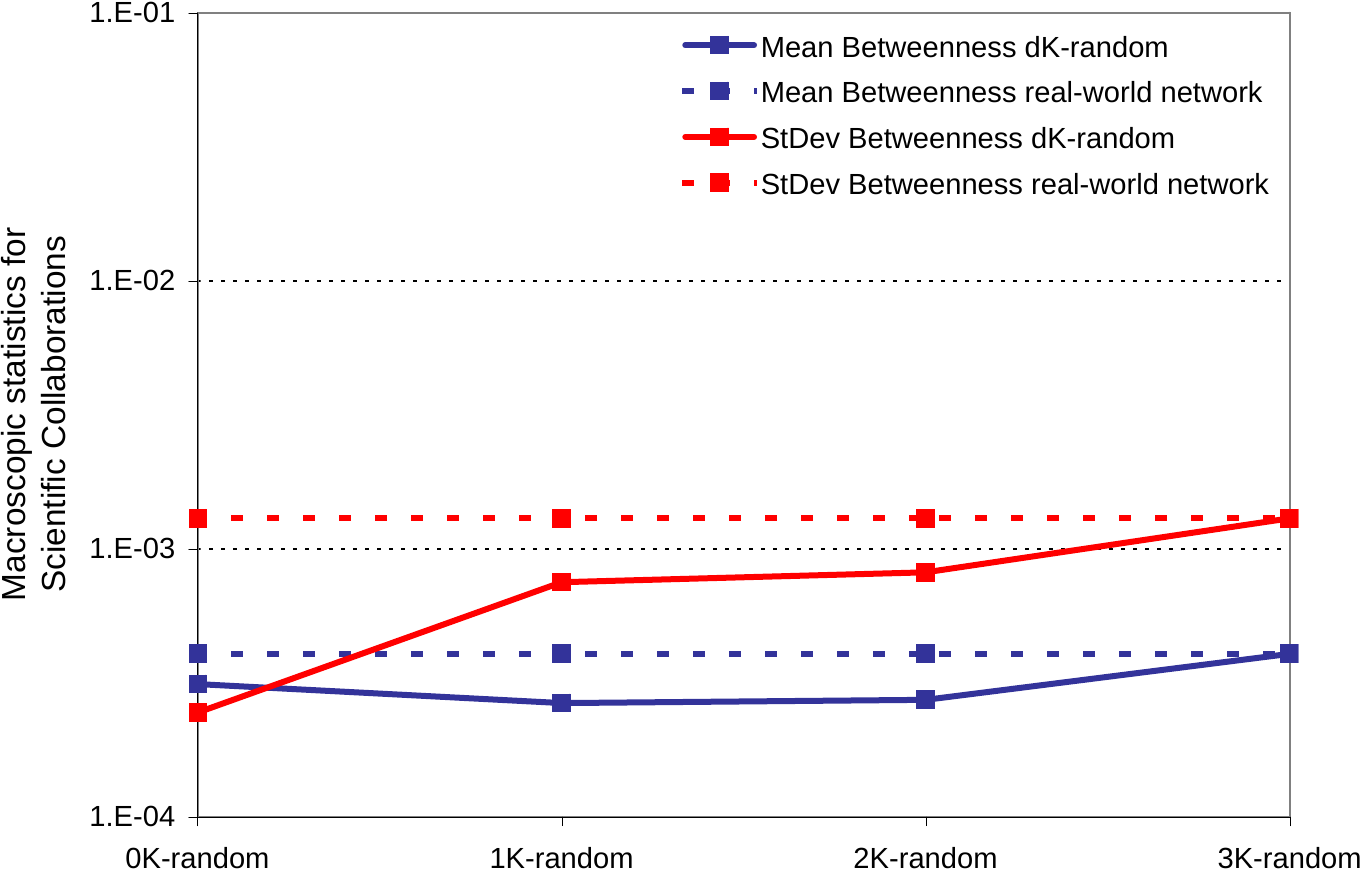}}
\subfigure{\includegraphics[width=2.25in]{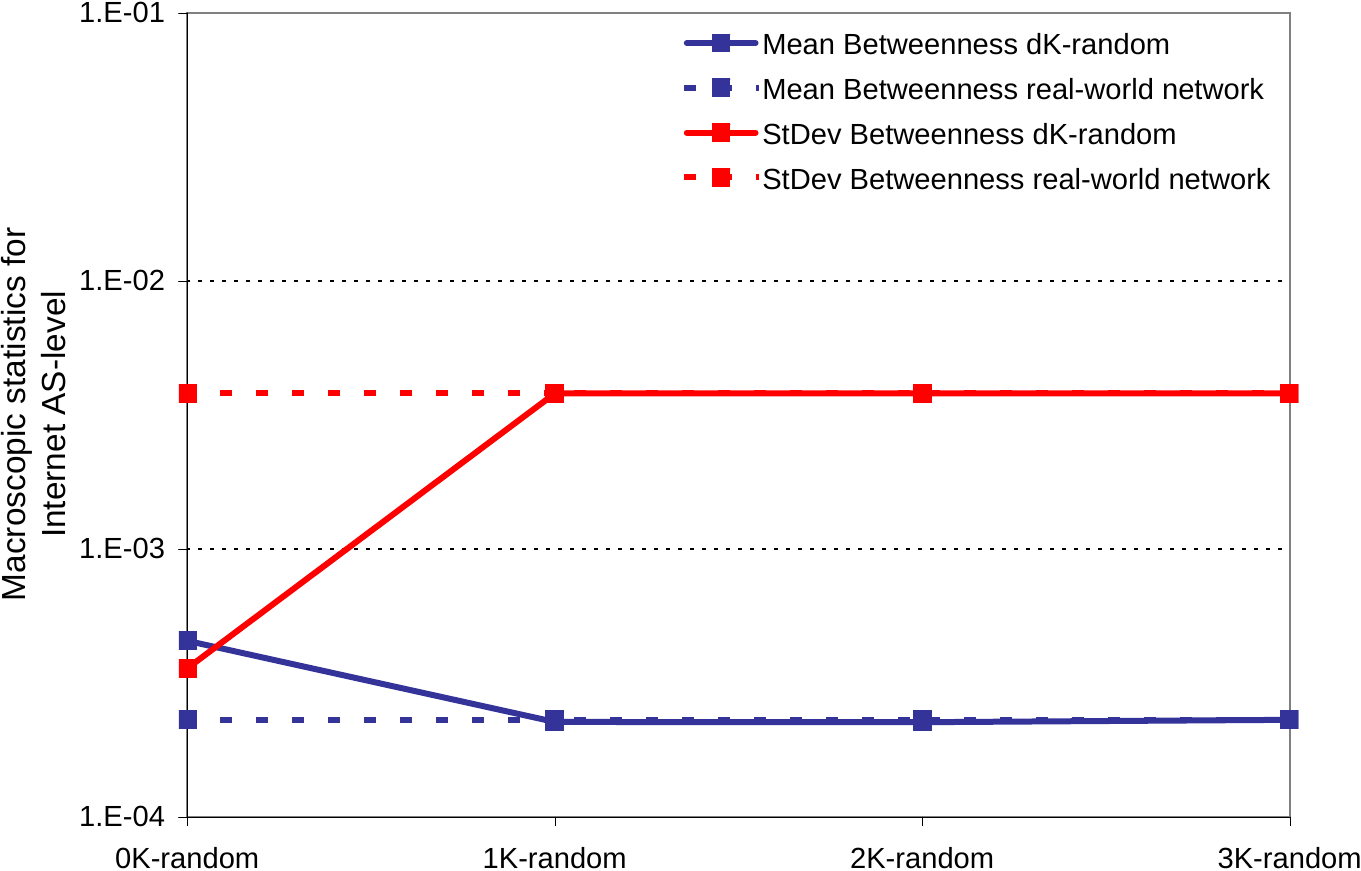}}\\
\subfigure{\includegraphics[width=2.25in]{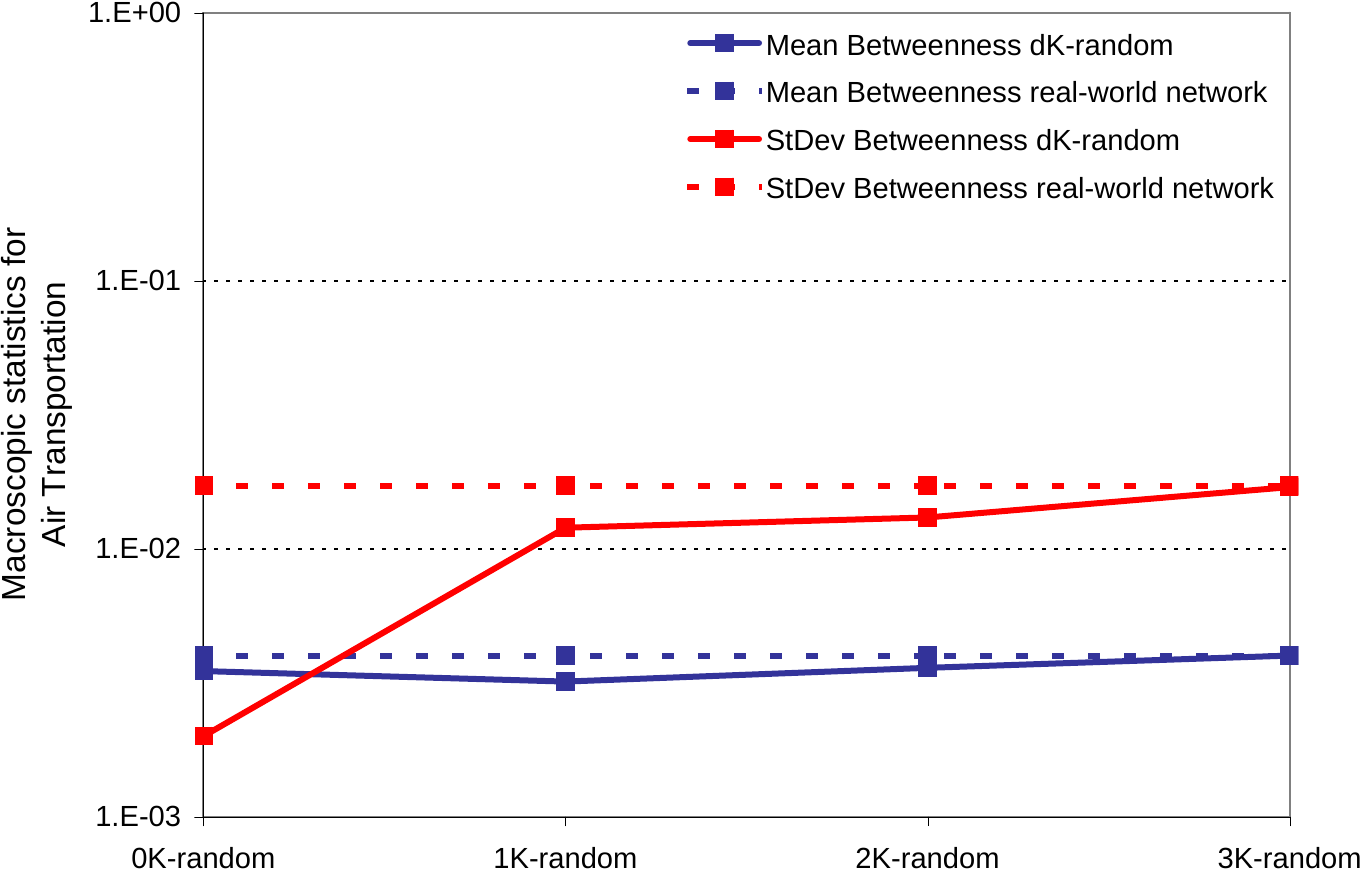}}
\subfigure{\includegraphics[width=2.25in]{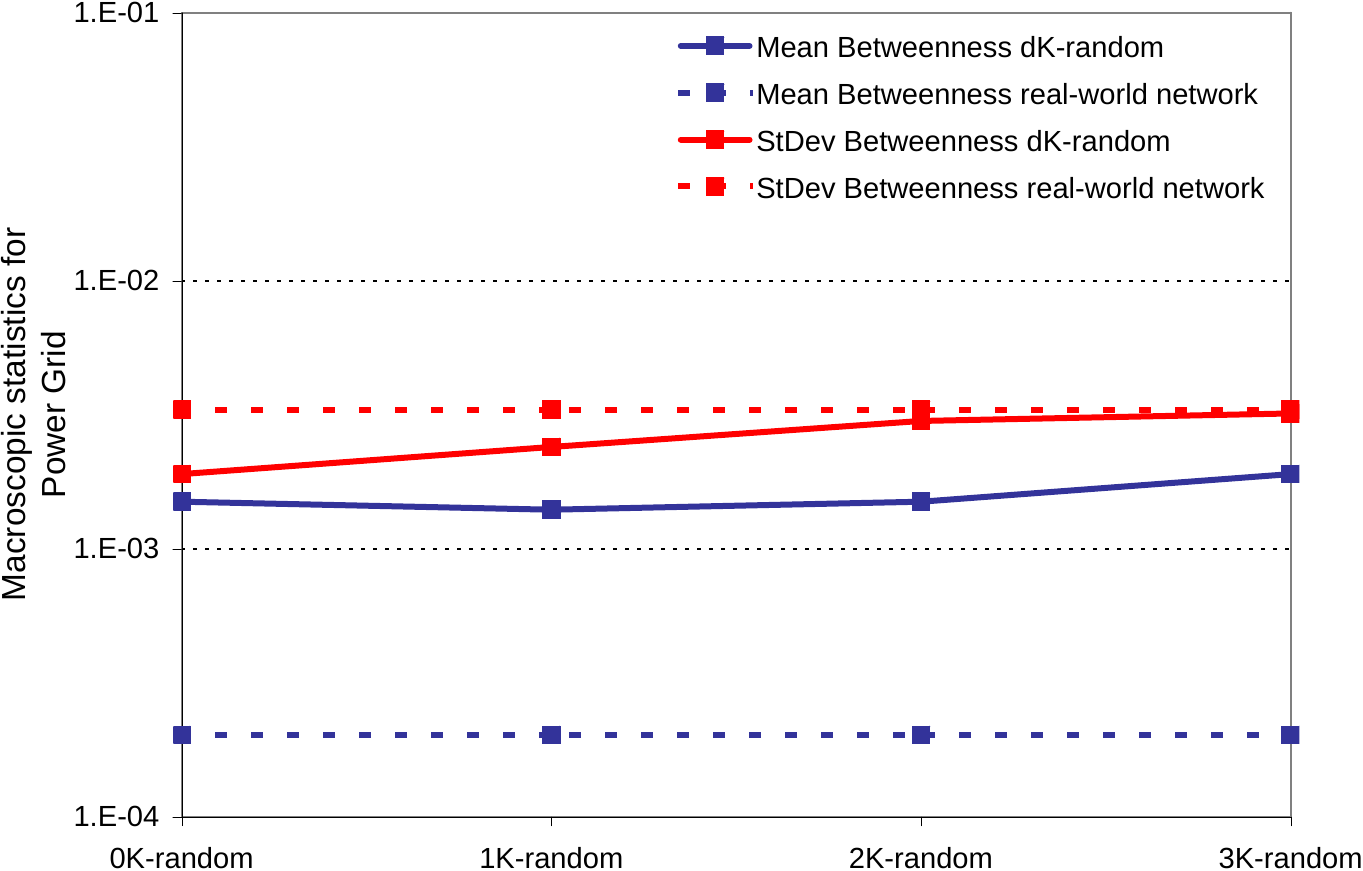}}
\caption{The average betweenness and the standard deviation of the
betweenness distribution as functions of $d$ for $dK$-randomisations
of the real networks. The corresponding values for the real networks
are shown by dashed lines.} \label{AverageBetweennessGivenDegree(d)}
\end{figure*}

Fig.~\ref{DistanceDistribution} shows the distance distribution in
the real networks and in their $dK$-randomizations. The distance
distribution is the distribution of hop-lengths of shortest paths
between nodes in a network. Formally, if $N(h)$ is the number of
node pairs located at hop distance $h$ from each other, then the
distance distribution $\delta(h)$ is
\begin{equation}
\delta(h) = \frac{2N(h)}{N(N-1)},
\end{equation}
where $N(N-1)/2$ is the total number of nodes pairs in the network.

To provide a clearer view of how close the distance distributions in
$dK$-randomizations are to the real networks, we show in
Fig.~\ref{DistanceDistribution(d)} some scalar summary statistics of
the distance distribution as functions of $d$. These summary
statistics are the average distance
\begin{equation}\label{eq:bar_h}
\bar{h} = \sum_hh\delta(h),
\end{equation}
and the standard deviation of the distance distribution $\delta(h)$.
In addition we show in Fig.~\ref{DistanceDistribution(d)} the
network diameter, i.e., the maximum hop-wise distance between nodes
in the network, which is an extremal statistics of the distance
distribution.

Fig.~\ref{AverageBetweennessGivenDegree} shows degree-dependent
betweenness centrality $\bar{b}(k)$ in the real networks and their
$dK$-randomizations. Betweenness $b(i)$ of node $i$ is a measure of
how ``important'' $i$ is in terms of the number of shortest paths
passing through it. Formally, if $\sigma_{st}(i)$ is the number of
shortest paths between nodes $s \neq i$ and $t \neq i$ that pass
through $i$, and $\sigma_{st}$ is the total number of shortest paths
between the two nodes $s \neq t$, then betweenness of $i$ is
\begin{equation}
b(i) = \sum_{s,t}\frac{\sigma_{s,t}(i)}{\sigma_{s,t}}.
\end{equation}
Averaging over all nodes of degree $k$, degree-dependent betweenness
$\bar{b}(k)$ is
\begin{equation}
\bar{b}(k) = \sum_{i:\,\deg(i)=k}\frac{b(i)}{N(k)}.
\end{equation}

We also compute the betweenness distribution, and show its average
and standard deviation in
Fig.~\ref{AverageBetweennessGivenDegree(d)}.

We observer similar trends with respect to both distance and
betweenness metrics. The power grid cannot be approximated even by
its $3K$-randomization. The Internet lies at the other extreme: even
$1K$-randomization does not disturb its global metrics too much. The
air transportation network appears to come next, as its
$2K$-randomizations resemble it closely. But all the networks other
than the power grid are very similar to their $3K$-randomizations.

\subsection{Scalar topological metrics and $dK$-randomness of real networks}

\begin{table}
\caption{The scalar topological metrics of the real networks and the
minimum value of $d$ such that the network's $dK$-randomizations
approximately preserve all the metrics.} \label{dk-table}
\begin{tabular}{|c||c|c|c|c|c|c|}
\hline Metrics & PGP & Collab. & Protein & Air & Internet & Power
\\ \hline
   \hline $\bar{k}$ & 4.6 &  6.4 & 6.4 & 11.9  &  6.3 &  4.7
\\ \hline $r$       & 0.238 &  0.157 &  -0.137 & -0.268 &  -0.236 &  -0.273
\\ \hline $\bar{c}$ & 0.27 &  0.65 &  0.09 & 0.62 &  0.46 &  0.68
\\ \hline $\bar{h}$ & 7.5 &  6.6 &  4.2 & 3.0 &  3.1 &  2.0
\\ \hline $\bar{b}$ & $6\cdot10^{-4}$ &  $4\cdot10^{-4}$ &  $7\cdot10^{-4}$ & $4\cdot10^{-3}$ &  $2\cdot10^{-4}$ &  $2\cdot10^{-4}$
\\ \hline $dK$      & $3K$ &  $3K$ &  $3K$ &  $2K$ &  $1K$ & ?
\\ \hline
\end{tabular}
\end{table}

To conclude this section we show in Table~\ref{dk-table} the most
important scalar topological metrics for the real networks. These
metrics are coarse summary statistics of the more informative and
detailed metrics that we have considered in this section.
Specifically, these coarse summaries are:
\begin{itemize}
\item
$\bar{k}$ is the average degree in the network,
Eq.~(\ref{eq:bar_k}), which is both the $0K$-distribution and a
summary statistics of the $1K$-distribution in the $dK$-series
terminology;
\item
$r$ is the assortativity coefficient,
\begin{equation}
r=\frac{\langle k \rangle^2 \displaystyle{\sum_{k k'}kk'P(k,k')}-\langle k^2 \rangle^2}{\langle k^3 \rangle \langle k \rangle-\langle k^2 \rangle^2}
\end{equation}
which is nothing but the Pearson correlation coefficient of the
$2K$-distribution $P(k,k')$;
\item
$\bar{c}$ is the average clustering
\begin{equation}
\bar{c} = \sum_k\bar{c}(k)P(k),
\end{equation}
which is a coarse summary statistics of the $3K$-distribution;
\item
$\bar{h}$ is the average distance, Eq.~(\ref{eq:bar_h}), which is
unrelated to $dK$-distributions;
\item
$\bar{b}$ is the average betweenness,
\begin{equation}
\bar{b} = \sum_k\bar{b}(k)P(k),
\end{equation}
unrelated to $dK$-distributions as well.
\end{itemize}
In Table~\ref{dk-table} we also show the minimum value of $d$ such
the $dK$-randomization null model approximately reproduces the real
network with respect to all the metrics above.

The observation that the power grid cannot be approximated even by
its $3K$-randomization is instructive. It shows that there are real
networks for which no $d\leq3$ is capable of
preserving the network structure upon $dK$-randomizing. In case of
the power grid, the explanation why this network is not even
$3K$-random may be related to the fact that it is designed
and controlled by human engineers in a single organization.
Informally, we can think of
it as rather ``non-random,'' designed, and thus bearing a number of
constraints that the $dK$-distributions with low $d$ cannot capture.
Informally, the higher $d$ required to approximately preserve the
network structure upon $dK$-randomization, the less ``random'' the
network is. The commonly referred explanation that the power grid is
an ``outlier'' because it is spatially embedded, may be less
relevant here because two other networks that we have considered
(the Internet and air transportation) are also spatially embedded.

What is different between the power grid and the other considered
networks is that the latter are self-evolving. They may be
engineered to a certain degree, such as the Internet, but their
global structure and evolution are not fully controlled by any
single human or organization. In the Internet case, for example, the
global network topology is a cumulative effect of independent
decisions made by tens of thousands of separate organizations,
roughly corresponding to Autonomous Systems, i.e., nodes of the
Internet graph.

In that sense, self-evolving complex networks are ``more random.''
However, why the level of their ``randomness'' is at $d\leq3$
remains an open question.

\section{Motif-based series vs.\ $dK$-series}\label{sec:motifs-vs-dk}

In this section we compare $dK$-series with the series based on
motifs, and show that the latter cannot form a systematic basis for
topology analysis.

The difference between $dK$-series and motif-series, which we can
call $d$-series, is that the former is the series of distributions
of $d$-sized subgraphs labeled with node degrees in a given network,
while the $d$-series is the distributions of such subgraphs in which
this degree information is ignored. This difference explains the
mnemonic names for these two series: `$d$' in `$dK$' refers to the
subgraph size, while `$K$' signifies that they are labeled by node
degrees---`$K$' is a standard notation for node degrees.

This difference between the $dK$-series and $d$-series is crucial.
The $dK$-series are inclusive, in the sense that the
$(d+1)K$-distribution contains the full information about the
$dK$-distribution, plus some additional information, which is not
true for $d$-series.

\begin{table}
\caption{$dK$-series vs.\ $d$-series} \label{dk.vs.d}
\begin{tabular}{|c||c|c|}
\hline $d$ & $dK$-statistics & $d$-statistics
\\ \hline
   \hline $0$ & $\bar{k}$ & -
\\ \hline $1$ & $N(k)$    & $N$
\\ \hline $2$ & $N(k,k')$ & $M$
\\ \hline \multirow{2}{*}{$3$} & $N_\wedge(k,k',k'')$    & $W$
\\                             & $N_\triangle(k,k',k'')$ & $T$
\\ \hline
\end{tabular}
\end{table}

To see this, let us consider the first few elements of both series
in Table~\ref{dk.vs.d}. In Section~\ref{sec:dk-metrics} we show
explicitly how the $(d+1)K$-distributions define the
$dK$-distribution for $d=0,1,2$. The key observation is that the
$d$-series does not have this property. The $0$'th element of
$d$-series is undefined. For $d=1$ we have the number of subgraphs
of size $1$, which is just $N$, the number of nodes in the network.
For $d=2$, the corresponding statistics is $M$, the number of links,
subgraphs of size $2$. Clearly, $M$ and $N$ are independent
statistics, and the former does not define the latter. For $d=3$,
the statistics are $W$ and $T$, the total number of wedges and
triangles, subgraphs of size $3$, in the network. These do not
define the previous element $M$ either. Indeed, consider the
following two networks of size
$N$---the chain and the star:\\
\centerline{\includegraphics[width=3in]{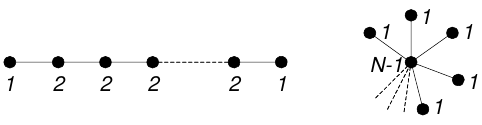}}
There are no triangles in either network, $T=0$. In the chain
network, the number of wedges is $W=N-2$, and in the
star $W=(N-1)(N-2)/2$. We see that even though $W$ ($d=3$) scales
completely differently with $N$ in the two networks, the number of
edges $M=N-1$ ($d=2$) is the same.

In summary, $d$-series is not inclusive. For each $d$, the
corresponding element of the series reflects a differen kind of
statistical information about the network topology, unrelated or
only loosely related to the information conveyed by the preceding
elements. At the same time, similar to $dK$-series, the $d$-series
is also converging since at $d=N$ it specifies the whole network
topology. However, this convergence is much slower that in the
$dK$-series case. In the two networks considered above, for example,
neither $W=N-2,\; T=0$ nor $W=(N-1)(N-2)/2,\; T=0$, fix the network
topology as there are many non-isomorphic graphs with the same
$(W,T)$ counts, whereas the $3K$-distributions $N_\wedge(1,2,2)=2,\;
N_\wedge(2,2,2)=N-4$ and $N_\wedge(1,N-1,1)=(N-1)(N-2)/2$ define the
chain and star topologies exactly.

The node degrees thus provide necessary information about subgraph
locations in the original network, which improves convergence, and
makes the $dK$-series basis inclusive and systematic.

\begin{acknowledgments}
We thank KC Claffy, Marina Fomenkov, Alex Arenas, and Alessandro Vespignani for useful comments
and discussions, and Connie Lyu and Bradley Huffaker for their help
with
Figs.~\ref{fig:dk-series},\ref{fig:dk-series-appendix}.
This work was supported in part by
DGES grant FIS2007-66485-C02-02, by NSF CNS-0434996 and CNS-0722070,
by DHS N66001-08-C-2029, and by Cisco Systems.
\end{acknowledgments}


%

\end{document}